\documentclass[fleqn,twoside,twocolumn,nofootinbib,showkeys]{revtex4} 
\usepackage[sec,nocpr]{ujp} 

\def\lsim{\
  \lower-1.2pt\vbox{\hbox{\rlap{$<$}\lower5pt\vbox{\hbox{$\sim$}}}}\ }
\def\gsim{\
  \lower-1.2pt\vbox{\hbox{\rlap{$>$}\lower5pt\vbox{\hbox{$\sim$}}}}\ }

\newcommand{\sss}{\scriptstyle}

\begin{document}
\title[Low-lying energy levels of a one-dimensional weakly interacting Bose gas
 under  zero boundary conditions]
{LOW-LYING ENERGY LEVELS \\ OF A ONE-DIMENSIONAL WEAKLY INTERACTING \\ BOSE GAS UNDER ZERO BOUNDARY CONDITIONS}%
\author{M.D.~Tomchenko}
\affiliation{Bogolyubov Institute for Theoretical Physics, Nat. Acad. of Sci. of Ukraine}
\address{14b, Metrolohichna Str., Kyiv 03143, Ukraine}
\email{mtomchenko@bitp.kiev.ua}

\razd{\secx}

\autorcol{M.D.~Tomchenko}

\setcounter{page}{1}%

\begin{abstract}
We diagonalize the second-quantized Hamiltonian of a one-dimensional
Bose gas with a nonpoint repulsive interatomic potential and zero
boundary conditions. At weak coupling the solutions for the
ground-state energy $E_{0}$ and the dispersion law $E(k)$ coincide
with the Bogoliubov solutions for a periodic system. In this case,
the single-particle density matrix $F_{1}(x,x^{\prime})$ at $T=0$ is
close to the solution for a periodic system and, at $T>0$, is
significantly different from it. We also obtain that the wave
function $\langle \hat{\psi}(x,t) \rangle$ of the effective
condensate is close to a constant $\sqrt{N_{0}/L}$ inside the system
and vanishes on the boundaries (here, $N_{0}$ is the number of atoms
in the effective condensate, and $L$ is the size of the system). We
find the criterion of applicability of the method, according to
which the method works for a finite system at very low temperature
and with a weak coupling (a weak interaction or a large
concentration).
\end{abstract}

 \maketitle

\section{Introduction}
In the description of uniform many-particle Bose systems, the real
boundary conditions (BCs) are usually replaced by periodic ones
\cite{bog1947,fey1954,bz1955,brueck1959,ll1963}, since it simplifies
the finding of a solution. However, a BCs in the Nature are usually
close to zero ones. Therefore, it is of significance to solve the
problem for zero BCs.  Such problem is interesting and not quite
trivial from the mathematical point of view. From the physical
viewpoint, it is very important whether the boundaries affect the
bulk properties of the system such as the ground-state energy
$E_{0}$, dispersion law $E(k)$, the condensate and thermodynamic
parameters. It is usually assumed that such influence is absent.
However, such influence is possible and, apparently, does not
contradict any physical laws \cite{zero-liquid}. The solutions for a
Bose system under zero BCs were found for a point interaction
\cite{gaudin1971,mt2015,mtjpa2018}. According to those results,
$E_{0}$ and $E(k)$ of a one-dimensional (1D) system under zero BCs
differ from $E_{0}$ and $E(k)$ of a system under periodic BCs only
by a negligible surface correction. For a nonpoint interaction, the
problem was solved in the Haldane's harmonic-fluid approach
\cite{cazalilla2002,cazalilla2004}. It was found that the sound
velocity in a system under zero BCs is the same as in a system under
periodic BCs \cite{cazalilla2004}. The attempt was made
\cite{zero-liquid} to determine $E_{0}$ and $E(k)$ beyond the
harmonic-fluid approximation. However, this calculation contains the
following weak place. The ground-state wave function of the system
was sought in the form $\Psi_{0} =
e^{S}\prod\limits_{j=1}^{N}\sin{(\pi x_{j}/L)}$ (for 1D), and then
the equation for $S$ was solved. In this case, zero BCs hold
automatically. Therefore, the differential equation for $S$ is not
accompanied by BCs and has a continuous set of solutions that
consists of the infinite number of discrete collections. One of
these discrete collections is the required solution of the problem.
But the probability to find just it is not too high: $1/\infty =0$.
In other words, the BCs were lost on a definite step of the method
\cite{zero-liquid}.

In the present work, a solution of the problem with zero BCs will be
found for arbitrary nonpoint repulsive potential, which can be
expanded in a Fourier series, on the basis of the Bogoliubov method
\cite{bog1947}. To simplify the calculations, the 1D case will be
considered. Though the Bogoliubov method was developed
\cite{girardeau1959,bogquasi,fetter1972,gardiner1997,girardeau1998,leggett2001,zagrebnov2001,zagrebnov2007},
we will be guided by the original approach \cite{bog1947}, since it
combines the simplicity and ability to describe the main properties
of a system.

Note that a modified Bogoliubov method was used for describing a
weakly interacting Bose gas in harmonic trap
\cite{rovenchak2007,rovenchak2016}. Moreover, the possibility to
describe a Bose liquid as the Coulomb system of electrons and nuclei
is discussed in \cite{bzt2015}.

The solution obtained below is valid for a large number of
particles: $N\gsim 1000$. The system with $N\lsim 100$ can be
described in the exactly solvable approach
\cite{ll1963,gaudin1971,mt2015,mtjpa2018,deguchi2013} (for a
point-like potential) or by the multiconfigurational time-dependent
Hartree method for bosons  \cite{streltsov2013} (for a potential of
the general form).

In Section 5 we obtain the criterion of applicability of the method.
According to it, the method can be used at low temperatures $T$,
small coupling constant $\gamma,$ and large finite $N$. In Section 6
we show that the single-particle density matrix
$F_{1}(x_{1},x_{1}+x)$ as a function of $x$ decreases, as $x$
increases: by a power law at $T=0$ and exponentially at $T> 0$. At
such properties the term ``quasicondensate'' is usually used instead
of the condensate (at the exponential decrease, the term
quasicondensate can be applied at low temperatures, for which
$F_{1}(x_{1},x_{1}+x)$ decreases weakly). In the region of
parameters for which the method is valid, the quasicondensate is
close to the true condensate. The key aspect consists in that the
system should be \textit{finite}. In an infinite 1D system, the
condensate (quasicondensate) is impossible for $T>0$
\cite{bogquasi,kk1967,hohenberg1967}. The difference between finite
and infinite 1D systems is easily seen from formulae in
\cite{bog1947,hohenberg1967}: for a finite system, one needs to
integrate over $|k|\in [2\pi/L,\infty]$ instead of $|k|\in
[0,\infty]$, which leads to a possibility for the condensate to
exist at $T>0$. The more strict analysis of the condensation of
atoms and quasiparticles in a finite system can be found,
respectively, in \cite{fischer2002} and \cite{bl2007}.

\section{Starting reasonings}
Consider $N$ spinless Bose particles that are located on the
interval $[0,L]$  and interact  via a repulsive potential of the
general form $U(|x_{j}-x_{l}|)$. Our study is essentially based  on
the classical work by Bogoliubov \cite{bog1947}, and we will try to
conserve its notations for the convenience of a reader. The
Hamiltonian  reads
\begin{equation}
 \hat{H}= -\frac{\hbar^2}{2m}\sum\limits_{j=1}^{N}\frac{\partial^2}{\partial x_{j}^2}+
 \sum\limits_{j<l}U(|x_{j}-x_{l}|),
     \label{1} \end{equation}
and zero BCs are as follows:
 \begin{equation}
  \hat{\psi}(0,t)  = 0, \quad \hat{\psi}(L,t)  = 0.
     \label{3} \end{equation}
Any operator $\hat{\psi}(x,t)$ satisfying Eqs. (\ref{3}) can be
expanded in the complete orthonormalized collection of sines:
\begin{equation}
  \hat{\psi}(x,t)  = \sum\limits_{j=1,2,\ldots,\infty}\hat{d}_{j}(t)
  \sqrt{2/L}\cdot   \sin(k_{j}x),
     \label{4} \end{equation}
where $k_{j}=\pi j/L$.  In the Bogoliubov work \cite{bog1947}, the
expansion $\hat{\psi}(x,t) =\frac{1}{\sqrt{L}}\sum_{j=0,\pm
1,\ldots,\pm\infty}\hat{d}_{j}(t)e^{i2k_{j}x}$ is used and it is
assumed that almost all atoms are in the state  $j=0$ (therefore,
the replacement $\hat{d}_{0}\rightarrow d_{0}$ is executed). This
allows one to construct the description of weakly excited states of
a uniform system with weak coupling.

For the \textit{nonuniform} system we should apply a more general
approach. There are available two different definitions of a
condensate for nonuniform systems. The first is grounded  on the
representation \cite{leggett2001,kirzhnits1978,griffin2002}
\begin{equation}
\hat{\psi}(x,t)  = \langle \hat{\psi}(x,t)\rangle
+\hat{\vartheta}(x,t).
     \label{7} \end{equation}
The system contains a condensate described by the wave function
$\langle \hat{\psi}(x,t)\rangle$, if $N_{0}\equiv\int_{0}^{L}
dx|\langle \hat{\psi}(x,t)\rangle|^{2}\sim N$. The second definition
\cite{leggett2001,penronz} is based on the diagonal expansion of the
single-particle density matrix $F_{1}(x,x^{\prime})$ in the complete
orthonormalized basis $\{\phi_{j}(x)\}$:
\begin{equation}
F_{1}(x,x^{\prime})  =
\sum\limits\limits_{j=0}^{\infty}\lambda_{j}\phi^{*}_{j}(x^{\prime})\phi_{j}(x).
     \label{n2} \end{equation}
If $\lambda_{0}\sim N$, then a condensate is present in the state
$\phi_{0}(x)$. From the physical point of view the primary criterion
is (\ref{n2}), since  $\lambda_{j}/N$ is equal to the probability
for a particle to be in the state $\phi_{j}(x)$. Whereas criterion
(\ref{7}) is rather a way to use a single-particle function in the
description of a many-particle system. These two criteria are
equivalent for a uniform periodic system (Bogoliubov solution
\cite{bog1947}). However, they can be \textit{nonequivalent} for a
nonuniform system \cite{frag2018}. Therefore, we will consider the
condensate defined on the basis of criterion (\ref{n2}) to be
genuine. In this case, we will call the order parameter $\langle
\hat{\psi}(x,t)\rangle$ the effective condensate.

Here and below, $\langle \rangle$ stands for the statistical average
\cite{bog1949}:
\begin{equation}
\langle \hat{A}\rangle_{T>0}=\frac{1}{Z}\int dx_{1}\ldots
dx_{N}\sum\limits_{p}e^{-E_{p}/k_{B}T}\Psi^{*}_{p}\hat{A}\Psi_{p},
      \label{srT} \end{equation}
where $Z=\sum_{l}e^{-E_{l}/k_{B}T}$, and
$\{\Psi_{p}(x_{1},\ldots,x_{N})\}$ is the complete orthonormalized
set of wave functions of a system with a fixed number of particles
$N$. For the pure state, which is possible in the many-particle
system only at $T=0$, it is the quantum-mechanical average:
\begin{equation}
\langle \hat{A}\rangle_{T=0}=\int dx_{1}\ldots
dx_{N}\Psi^{*}_{0}\hat{A}\Psi_{0}.
     \label{sr0} \end{equation}
The operator $\hat{\psi}(x,t)$ can be expanded in any
single-particle basis satisfying zero BC. In this case, the operator
$\hat{\psi}(x,t)$, values of $E_{p},$  functions $\Psi_{p}$, and the
function $\langle \hat{\psi}(x,t)\rangle$  are independent of the
basis.

Relations (\ref{4}) and (\ref{7}) yield
\begin{equation}
\hat{d}_{j}(t)=\langle\hat{d}_{j}(t)\rangle+\hat{a}_{j}(t)\equiv
d_{j}(t)+\hat{a}_{j}(t),
  \label{9b} \end{equation}
\begin{equation}
 \langle \hat{\psi}(x,t)\rangle=
\sum_{j=1,2,\ldots,\infty}d_{j}(t)\sqrt{2/L}\cdot\sin(k_{j}x),
      \label{8-0}    \end{equation}
\begin{equation}
\hat{\vartheta}(x,t)  =
\sum_{l=1,2,\ldots,\infty}\hat{a}_{l}(t)\sqrt{2/L}\cdot\sin(k_{l}x),
     \label{9} \end{equation}
\begin{eqnarray}
&&\hat{\psi}^{+}(x,t)=\sum_{j=1,2,\ldots,\infty}\hat{d}^{+}_{j}(t)\sqrt{2/L}\cdot\sin(k_{j}x)=%
\nonumber \\ &&= \langle \hat{\psi}^{+}(x,t)\rangle +%
\hat{\vartheta}^{+}(x,t),%
     \label{10} \end{eqnarray}
\begin{equation}
\hat{\vartheta}^{+}(x,t)  =
\sum_{l=1,2,\ldots,\infty}\hat{a}^{+}_{l}(t)\sqrt{2/L}\cdot\sin(k_{l}x).
     \label{11} \end{equation}
The Bose operators $\hat{d}^{+}_{l}$ and $\hat{d}_{j}$ satisfy the
commutation relations
\begin{equation}
\hat{d}_{j}\hat{d}^{+}_{l}-\hat{d}_{l}^{+}\hat{d}_{j}=\delta_{l,j},
\ \
\hat{d}_{j}\hat{d}_{l}-\hat{d}_{l}\hat{d}_{j}=\hat{d}_{j}^{+}\hat{d}^{+}_{l}-\hat{d}_{l}^{+}\hat{d}_{j}^{+}=0.
     \label{12} \end{equation}
The operators $\hat{a}^{+}_{l}$ and
$\hat{a}_{j}$ satisfy the same relations.

The subsequent analysis is based on relations (\ref{7}) and
$\hat{\vartheta}\ll \langle \hat{\psi}(x,t)\rangle$. They are basic
formulae allowing one to construct the description of a nonuniform
weakly interacting Bose gas.

\section{Method 1: solving the operator equation}
The Heisenberg equation for the operator $\hat{\psi}=\langle
\hat{\psi}(x,t)\rangle + \hat{\vartheta}$ reads
\begin{eqnarray}
&& i\hbar\frac{\partial \hat{\psi}(x,t)}{\partial t}  =
-\frac{\hbar^2}{2m}\frac{\partial^{2} \hat{\psi}(x,t)}{\partial
x^{2}}
 + \nonumber \\ &&+ \int\limits_{0}^{L} dx^{\prime} U(|x-x^{\prime}|)\hat{\psi}^{+}(x^{\prime},t)\hat{\psi}(x^{\prime},t)\hat{\psi}(x,t).
     \label{2} \end{eqnarray}
At $\hat{\vartheta}\ll \langle \hat{\psi}(x,t)\rangle$ it can be
separated into two equations (in this Section, we denote $\langle
\hat{\psi}(x,t)\rangle\equiv\psi_{0}(x,t)$):
\begin{eqnarray}
&& i\hbar\frac{\partial \psi_{0}(x,t)}{\partial t}  =
-\frac{\hbar^2}{2m}\frac{\partial^{2} \psi_{0}(x,t)}{\partial x^{2}}
+ \nonumber \\ &&
 + \psi_{0}(x,t)\int\limits_{0}^{L} dx^{\prime} U(|x-x^{\prime}|)|\psi_{0}(x^{\prime},t)|^{2},
     \label{13} \end{eqnarray}
\begin{eqnarray}
 && i\hbar\frac{\partial \hat{\vartheta}(x,t)}{\partial t}  = -\frac{\hbar^2}{2m}\frac{\partial^{2} \hat{\vartheta}(x,t)}{\partial x^{2}}
 + \nonumber \\ && + \hat{\vartheta}(x,t)\int\limits_{0}^{L} dx^{\prime} U(|x-x^{\prime}|)|\psi_{0}(x^{\prime},t)|^{2}+ \label{14} \\
 &&+ \psi_{0}(x,t)\int\limits_{0}^{L} dx^{\prime} U(|x-x^{\prime}|)\left [\psi_{0}(x^{\prime},t)
  \hat{\vartheta}^{+}(x^{\prime},t)+\right. \nonumber \\
 &&+\left. \psi^{*}_{0}(x^{\prime},t)\hat{\vartheta}(x^{\prime},t)\right ].
     \nonumber \end{eqnarray}
Eq.(\ref{13}) was first obtained by Gross \cite{gross1957} and is
usually called the Gross--Pitaevskii equation
\cite{gross1961,pit1961}. It is clear that the lowest energy
solution of Eq. (\ref{13}) is given by the function
$\psi_{0}(x,t)=\psi_{0}(t)$. To satisfy zero BCs, $\psi_{0}(x,t)$
must decrease to zero near the boundaries. Neglecting this
nonuniformity, we can write $\psi_{0}(x,t)=a_{0}(t)/\sqrt{L}$ (more
accurate solution for $\langle \hat{\psi}(x,t)\rangle$ will be
obtained in the next Section). Set $a_{0}^{*}a_{0}=N_{0}$ and
$n_{0}=N_{0}/L$. Then Eq. (\ref{13}) takes the form
\begin{equation}
 i\hbar\frac{\partial a_{0}(t)}{\partial t}  = a_{0}(t)n_{0}\int\limits_{0}^{L} dx^{\prime} U(|x-x^{\prime}|).
     \label{15} \end{equation}
We expand the potential in the Fourier series:
\begin{equation}
 U(|x_{1}-x_{2}|) = \frac{1}{2L}
 \sum\limits_{ j=0,\pm 1,\pm 2,
 \ldots}\nu(k_{j})e^{ik_{j}(x_{1}-x_{2})},
     \label{16} \end{equation}
\begin{equation}
 \nu(k_{j}) = \int\limits_{-L}^{L} U(|x|)e^{-ik_{j}x}dx, \quad k_{j}=\pi
 j/L.
     \label{17} \end{equation}
This series restores the potential exactly in the whole region
$x_{1},x_{2}\in [0,L]$ (expansions of the potential for different
BCs have been considered in \cite{mtpot2014}). Since $\nu(k_{j}) =
2\int\limits_{0}^{L} U(x)\cos(k_{j}x)dx=\nu(-k_{j})$, we have
\begin{eqnarray}
&& U(|x-x^{\prime}|) = \frac{\nu(0)}{2L}+ \sum\limits_{ j= 1, 2,
 \ldots}\frac{\nu(k_{j})}{L}\cdot\nonumber \\
 &&\cdot \left [\cos(k_{j}x)\cos(k_{j}x^{\prime})+
 \sin(k_{j}x)\sin(k_{j}x^{\prime})\right ].
     \label{18} \end{eqnarray}
Substituting expansion (\ref{18}) in (\ref{15}) and making some
transformations, we get the equation
\begin{equation}
 i\hbar\frac{\partial a_{0}(t)}{\partial t}  = a_{0}(t)n_{0}\nu(0)\left (1+S_{1}(x)\right ).
     \label{19} \end{equation}
The function $S_{1}(x)$  and the functions $S_{j}(x)$ arising below
are given and calculated in Appendix B. If the interaction radius
$a$ is small as compared with the system size  $L$, then all
functions $S_{j}(x)$ are negligibly small. We find  the solution of
Eq. (\ref{19}):
\begin{equation}
  a_{0}(t)=e^{\epsilon_{0}t/i\hbar}b_{0}, \quad \epsilon_{0}= n_{0}\nu(0).
     \label{20} \end{equation}
Taking into account zero BCs, the solution for $\psi_{0}$ can be
written in the form (\ref{8-0}):
\begin{eqnarray}
&& \psi_{0}(x,t)=
\frac{a_{0}(t)}{\sqrt{L}}\sum_{j=0,1,\ldots,\infty}\frac{4\sin(k_{2j+1}x)}{\pi
(2j+1)}
 =\nonumber \\&& =
\left [ \begin{array}{ccc}
    \frac{a_{0}(t)}{\sqrt{L}},  & \   \mbox{if} \ x\in ]0,L[,   & \\
    0,  & \mbox{if} \ x=0; L. &
\label{8} \end{array}           \right.            \end{eqnarray}

Now, let us substitute $\hat{\vartheta}$ (\ref{9}) and
$\hat{\vartheta}^{+}$ (\ref{11}) in Eq. (\ref{14}). With regard for
relations (\ref{18}) and (\ref{8}), Eq. (\ref{14}) is separated into
two equations: for the harmonics
 $\hat{a}_{2l}$ and $\hat{a}_{2j+1}$. The equation for $\hat{a}_{2l}$ reads
\begin{eqnarray}
&&i\hbar\sum\limits_{l=1,2,\ldots,\infty}\sin(k_{2l}x)\frac{\partial
\hat{a}_{2l}(t)}{\partial
t}=\sum\limits_{l=1,2,\ldots,\infty}\sin(k_{2l}x)\times \nonumber
\\ &&\times \hat{a}_{2l}(t)\left (K(k_{2l})+n_{0}\nu(0)[1+S_{1}(x)]\right
)+\nonumber \\
&&+ \sum\limits_{l=1,2,\ldots,\infty} \left
(n_{0}\hat{a}_{2l}(t)+\frac{a_{0}^{2}}{L}\hat{a}^{+}_{2l}(t)\right
)\times\nonumber \\&&\times \left
(\frac{\nu(k_{2l})}{2}\sin(k_{2l}x)+I_{1}(x,l)\right ),
     \label{21} \end{eqnarray}
\begin{equation}
I_{1}(x,l)=\sum\limits_{j=1,3,5,\ldots}\frac{\nu(k_{j})}{\pi}\cos(k_{j}x)\left
(\frac{1}{2l-j}+\frac{1}{2l+j}\right ),
     \label{21b} \end{equation}
where  $K(k)=\frac{\hbar^{2}k^{2}}{2m}$.  Using the expansion
\begin{equation}
\cos(k_{j}x)=\sum\limits_{p=1,2,3,\ldots}c_{j}^{p}\sin(k_{p}x),
\label{21d} \end{equation}
\begin{equation} c_{j}^{p} = \left [
\begin{array}{ccc}
0  & \   \mbox{for even} \ p-j,   & \\
\frac{2}{\pi}\left (\frac{1}{p-j}+\frac{1}{p+j}\right )  & \mbox{for
odd} \ p-j, & \end{array} \right.   \nonumber      \end{equation}
the sum $I_{1}$ (\ref{21b}) can be represented in the form
\begin{eqnarray}
&&I_{1}(x,l)=\nu(k_{2l})\sin(k_{2l}x)\left
(\frac{1}{2}+S_{2}(l)\right )+\nonumber
\\ &&+  \frac{2}{\pi^{2}}\sum\limits_{j=1,3,5,\ldots}\nu(k_{j})\sum\limits_{p=1,2,3,\ldots}^{p\neq l}\sin(k_{2p}x)\cdot
\nonumber \\ &&\cdot\left (\frac{1}{2l-j}+\frac{1}{2l+j}\right
)\left (\frac{1}{2p-j}+\frac{1}{2p+j}\right ).
     \label{21e} \end{eqnarray}
Then Eq. (\ref{21}) takes the form
\begin{eqnarray}
&&i\hbar\sum\limits_{l=1,2,\ldots,\infty}\sin(k_{2l}x)\frac{\partial
\hat{a}_{2l}(t)}{\partial t}=\nonumber
\\ &&=\sum\limits_{l=1,2,\ldots,\infty}\sin(k_{2l}x)
\frac{a_{0}^{2}}{L}\nu(k_{2l})\hat{a}^{+}_{2l}(t)(1+S_{2}(l)) +\nonumber \\
&&+\sum\limits_{l=1,2,\ldots,\infty}\sin(k_{2l}x)
\hat{a}_{2l}(t)\left [K(k_{2l})+\right. \label{21j}
\\ &&+\left. n_{0}\nu(0)(1+S_{1}(x))+n_{0}\nu(k_{2l})(1+S_{2}(l))\right
] + \nonumber \\
&&+ \sum\limits_{l=1,2,\ldots,\infty}
\sin(k_{2l}x)\sum\limits_{p=1,2,\ldots,\infty}^{p\neq l}\left (
n_{0}\hat{a}_{2p}(t)+\right. \nonumber \\ &&+\left.
\frac{a_{0}^{2}}{L}\hat{a}^{+}_{2p}(t)\right )S_{3}(l,p).
     \nonumber \end{eqnarray}
Since the functions $\sin(k_{2l}x)$ are independent, and the
corrections $S_{1}, S_{2},$ $ S_{3}$ are vanishingly small (see
Appendix B), Eq. (\ref{21j}) yields the system of equations
\begin{eqnarray}
&&i\hbar\frac{\partial \hat{a}_{2l}(t)}{\partial t}=\left
[K(k_{2l})+n_{0}\nu(0)+n_{0}\nu(k_{2l})\right
]\hat{a}_{2l}(t)+\nonumber
\\ &&+\frac{a_{0}^{2}\nu(k_{2l})}{L}\hat{a}^{+}_{2l}(t), \quad
l=1,2,\ldots,\infty.
     \label{23} \end{eqnarray}
In a similar way we find  the equation for the operators
$\hat{a}_{2j+1}$ from  (\ref{14}):
\begin{eqnarray}
&&i\hbar\sum\limits_{j=0,1,\ldots,\infty}\sin(k_{2j+1}x)\frac{\partial
\hat{a}_{2j+1}(t)}{\partial t}=\nonumber
\\ && = \sum\limits_{j=0,1,\ldots,\infty}\sin(k_{2j+1}x)
\frac{a_{0}^{2}}{L}\nu(k_{2j+1})\times \nonumber \\ &&\times \hat{a}^{+}_{2j+1}(t)(1+S_{4}(j))+  \label{22} \\
&&+\sum\limits_{j=0,1,\ldots,\infty}\sin(k_{2j+1}x)
\hat{a}_{2j+1}(t)\left [K(k_{2j+1})+\right. \nonumber
\\ && + \left. n_{0}\nu(0)(1+S_{1}(x))+n_{0}\nu(k_{2j+1})(1+S_{4}(j))\right
] + \nonumber \\
&&+ \sum\limits_{j=0,1,\ldots,\infty} \sin(k_{2j+1}x)\times\nonumber
\\&&\times \sum\limits_{l=0,1,\ldots,\infty}^{l\neq j}\left
(n_{0}\hat{a}_{2l+1}(t)+\frac{a_{0}^{2}}{L}\hat{a}^{+}_{2l+1}(t)\right
)S_{5}(l,j).
     \nonumber \end{eqnarray}
Since the corrections $S_{1}, S_{4},$ $S_{5}$ are small, and the
functions $\sin(k_{2j+1}x)$ are independent, Eq. (\ref{22}) leads to
the equations
\begin{eqnarray}
&&i\hbar\frac{\partial \hat{a}_{2j+1}(t)}{\partial t}=
\frac{a_{0}^{2}\nu(k_{2j+1})}{L}\hat{a}^{+}_{2j+1}(t)+\nonumber \\
&&+\left [K(k_{2j+1})+n_{0}\nu(0)+n_{0}\nu(k_{2j+1})\right
]\hat{a}_{2j+1}(t),
     \label{24} \end{eqnarray}
where $j=0,1,\ldots,\infty$.

 Thus, the starting equation (\ref{2}) is reduced to
Eqs. (\ref{19}), (\ref{23}), and (\ref{24}). Equations (\ref{23})
and (\ref{24}) coincide with the equation for $\hat{a}_{p}$ from
Bogoliubov work \cite{bog1947} with the only difference that our
equations involve $\hat{a}^{+}_{p}$ instead of $\hat{a}^{+}_{-p}$ in
\cite{bog1947}. This allows us to write the solution immediately.
Let us set $\hat{a}_{p}(t)=e^{\epsilon_{0}t/i\hbar}\hat{b}_{p}$
($p=1,2,\ldots,\infty$) and
$a_{0}(t)=e^{\epsilon_{0}t/i\hbar}b_{0}.$ With the help of
Bogoliubov transformations
\begin{equation}
\hat{b}_{p}=\frac{\hat{\xi}_{p}+\Lambda_{p}\hat{\xi}^{+}_{p}}{\sqrt{1-|\Lambda_{p}|^{2}}},
 \quad
\hat{b}^{+}_{p}=\frac{\hat{\xi}^{+}_{p}+\Lambda^{*}_{p}\hat{\xi}_{p}}{\sqrt{1-|\Lambda_{p}|^{2}}},
     \label{25a} \end{equation}
\begin{equation}
\hat{\xi}_{p}=\frac{\hat{b}_{p}-\Lambda_{p}\hat{b}^{+}_{p}}{\sqrt{1-|\Lambda_{p}|^{2}}},
 \quad
\hat{\xi}^{+}_{p}=\frac{\hat{b}^{+}_{p}-\Lambda^{*}_{p}\hat{b}_{p}}{\sqrt{1-|\Lambda_{p}|^{2}}},
     \label{25} \end{equation}
\begin{equation}
\Lambda_{p}=\frac{b_{0}^{2}}{N_{0}n_{0}\nu(k_{p})}\left
(E_{b}(k_{p})-K(k_{p})-n_{0}\nu(k_{p})\right ),
      \label{26} \end{equation}
we obtain  the following equations from (\ref{23}),  (\ref{24}):
\begin{eqnarray}
i\hbar\frac{\partial \hat{\xi}_{p}}{\partial
t}=E_{b}(k_{p})\hat{\xi}_{p}, \quad -i\hbar\frac{\partial
\hat{\xi}^{+}_{p}}{\partial t}=E_{b}(k_{p})\hat{\xi}^{+}_{p},
     \label{27} \end{eqnarray}
where $p=1,2,\ldots,\infty$ and
\begin{equation}
E_{b}(k)=\sqrt{K^{2}(k)+2n_{0}\nu(k)K(k)}.
      \label{28} \end{equation}
Our solution coincides with the Bogoliubov one \cite{bog1947} except
for the following difference: in Bogoliubov formulae (\ref{27}) $p$
runs the values $\pm 1, \pm 2, \ldots,\pm\infty$; whereas our $p$
are positive ($p=1,2,\ldots,\infty$), since the excitations are
standing waves. Formulae (\ref{27}) show that the excited state of
the system can be considered as the ideal gas of elementary
excitations with the dispersion law $E_{b}(k)$ (\ref{28}).

 \section{Method 2: diagonalization of the Hamiltonian}
In the second-quantized formalism Hamiltonian (\ref{1}) takes the
form \cite{bog1949}
\begin{equation}
\hat{H}=\hat{H}_{kin}+\hat{H}_{pot},
      \label{3-1} \end{equation}
\begin{eqnarray}
 \hat{H}_{kin}= -\frac{\hbar^2}{2m}\int\limits_{0}^{L}dx\hat{\psi}^{+}(x,t)\frac{\partial^2}{\partial
 x^2}\hat{\psi}(x,t),
     \label{3-2} \end{eqnarray}
\begin{eqnarray}
&&\hat{H}_{pot}= \frac{1}{2}\int\limits_{0}^{L}dx
dx^{\prime}U(|x-x^{\prime}|)
\hat{\psi}^{+}(x,t)\hat{\psi}^{+}(x^{\prime},t)\cdot\nonumber \\
&&\cdot\hat{\psi}(x,t)\hat{\psi}(x^{\prime},t).
     \label{3-3} \end{eqnarray}
We set $\hat{\psi}(x,t)  = \langle \hat{\psi}(x,t)\rangle
+\hat{\vartheta}(x,t)$, where $\langle \hat{\psi}(x,t)\rangle =
\psi_{0}(x,t)+\delta \psi_{0}(x,t)$. Here,
$\psi_{0}(x,t)=\sum_{j=1,2,\ldots,\infty}d^{(0)}_{j}(t)\sqrt{2/L}\cdot\sin(k_{j}x)$
is the bare part of the function $\langle \hat{\psi}(x,t)\rangle$
and is given by formula (\ref{8}), and $\delta \psi_{0}(x,t)$ is an
unknown small correction. First, we set $\delta \psi_{0}(x,t)=0$.
Then $\delta \psi_{0}(x,t)$ will be determined from the analysis. In
this case, while calculating $(\partial^{2}/\partial
 x^{2})\hat{\psi}(x),$ we should  take into account that the function
$\psi_{0}(x)$ (\ref{8}) is constant for $x\in ]0,L[$ and decreases
to zero by jump at the points $x=0,L$. Instead, it is simpler to use
the initial exact expansion (\ref{4}). In this case, we note that
(\ref{8}) yields
\begin{equation}
d^{(0)}_{2j}=0, \quad j=1,2,\ldots,\infty.
      \label{9b2} \end{equation}
 In such way with regard for the relations $\hat{d}_{j}(t)=
d^{(0)}_{j}(t)+\hat{a}_{j}(t)$,
$\hat{a}_{j}(t)=e^{\epsilon_{0}t/i\hbar}\hat{b}_{j}$, and
(\ref{9b2}), we obtain
\begin{eqnarray}
&&
\hat{H}_{kin}=\sum\limits_{l=1,2,\ldots,\infty}K(k_{2l})\hat{b}^{+}_{2l}\hat{b}_{2l}
+\nonumber
\\ &&+\sum\limits_{j=0,1,\ldots,\infty}K(k_{2j+1})(\hat{b}^{+}_{2j+1}+f^{*}_{2j+1})\cdot
\nonumber \\ && \cdot(\hat{b}_{2j+1} + f_{2j+1}),
     \label{3-4} \end{eqnarray}
where
\begin{equation}
f_{2j+1}=d^{(0)}_{2j+1}e^{-\epsilon_{0}t/i\hbar}=\frac{2\sqrt{2}b_{0}}{k_{2j+1}L}.
      \label{3-4b} \end{equation}

The operator $\hat{H}_{pot}$ contains no derivatives. Therefore, we
use the representation $\hat{\psi} = \psi_{0} + \hat{\vartheta}$
with $\psi_{0}$ (\ref{8}). Similarly to Bogoliubov work, we neglect
the terms of the 3-rd and 4-th degrees in $\hat{\vartheta}$ and
$\hat{\vartheta}^{+}$ in the product
$\hat{\psi}^{+}(x)\hat{\psi}^{+}(x^{\prime})\hat{\psi}(x)\hat{\psi}(x^{\prime})$.
Then
\begin{eqnarray}
 &&\hat{\psi}^{+}(x)\hat{\psi}^{+}(x^{\prime})\hat{\psi}(x)\hat{\psi}(x^{\prime})=\frac{(a_{0}a_{0}^{*})^{2}}{L^2}|_{1}+
\nonumber \\
 &&+ \frac{a_{0}^{*}a^{2}_{0}}{L^{3/2}}(\hat{\vartheta}^{+}(x)+\hat{\vartheta}^{+}(x^{\prime}))|_{3}
 +\frac{(a_{0}^{*})^{2}}{L}\hat{\vartheta}(x)\hat{\vartheta}(x^{\prime})|_{4}+\nonumber
 \\ && +\frac{a_{0}(a_{0}^{*})^{2}}{L^{3/2}}(\hat{\vartheta}(x)+\hat{\vartheta}(x^{\prime}))|_{2}
 +\frac{a_{0}^{2}}{L}\hat{\vartheta}^{+}(x)\hat{\vartheta}^{+}(x^{\prime})|_{5}+\nonumber \\
 &&+ \frac{a_{0}a_{0}^{*}}{L}\left
 (\hat{\vartheta}^{+}(x)+\hat{\vartheta}^{+}(x^{\prime})\right )
 \left (\hat{\vartheta}(x)+\hat{\vartheta}(x^{\prime})\right )|_{6}\equiv
 \nonumber \\ &&\equiv\sum\limits_{l=1,2,\ldots,6}\hat{\eta}_{l}.
     \label{3-5} \end{eqnarray}
Expression (\ref{3-5}) is a sum of six terms. Each of them is enumerated by the mark $|_{l}$ and is denoted by $\hat{\eta}_{l}$.
Using expansion (\ref{18}) of the potential, we get
\begin{eqnarray}
 \hat{H}_{pot}=\sum\limits_{l=1,2,\ldots,6}\hat{I}_{l},
     \label{3-6} \end{eqnarray}
\begin{eqnarray}
&&\hat{I}_{l}= \frac{1}{2}\int\limits_{0}^{L}dx
dx^{\prime}\hat{\eta}_{l}\left \{\frac{\nu(0)}{2L}+ \sum\limits_{ j=
1, 2, \ldots}\frac{\nu(k_{j})}{L}\times\right.\nonumber \\
&&\times\left.\left [\cos(k_{j}x)\cos(k_{j}x^{\prime})+
 \sin(k_{j}x)\sin(k_{j}x^{\prime})\right ]\right \}.
     \label{3-7} \end{eqnarray}
The calculation of integrals (\ref{3-7}) gives rather awkward sums,
and the main problem consists in the separation of their principal
parts. With regard for formulae  (\ref{8-0})--(\ref{11}),
(\ref{3-4b}), and $a_{0}(t)=e^{\epsilon_{0}t/i\hbar}b_{0}$,
$b_{0}^{*}b_{0}=N_{0}$, we obtain
\begin{eqnarray}
 \hat{I}_{1}=\frac{N_{0}n_{0}\nu(0)}{2}(1+S_{21}),
     \label{3-8} \end{eqnarray}
\begin{eqnarray}
 \hat{I}_{2}=\frac{n_{0}}{2}\sum\limits_{j=0,1,\ldots,\infty}f_{2j+1}^{*}\hat{b}_{2j+1}
 \left (\nu(0)+\nu(k_{2j+1}) \right ),
     \label{3-9} \end{eqnarray}
 \begin{eqnarray}
 \hat{I}_{3}=\frac{n_{0}}{2}\sum\limits_{j=0,1,\ldots,\infty}f_{2j+1}\hat{b}^{+}_{2j+1}
 \left (\nu(0)+\nu(k_{2j+1}) \right ),
     \label{3-9b} \end{eqnarray}
\begin{eqnarray}
 &&\hat{I}_{4}=\frac{(b_{0}^{*})^{2}}{2L}\left
 [\sum\limits_{l=1,2,\ldots,\infty}\nu(k_{2l})\hat{b}^{2}_{2l}(1+S_{22}(l))
 + \right.\nonumber \\&&+
 \sum\limits_{j=0,1,\ldots,\infty}\nu(k_{2j+1})\hat{b}^{2}_{2j+1}(1+S_{23}(j))+\nonumber
 \\ &&+
 \sum\limits_{l_{1},l_{2}=1,2,\ldots,\infty}^{l_{1}\neq
 l_{2}}\hat{b}_{2l_{1}}\hat{b}_{2l_{2}}S_{24}(l_{1},l_{2})+
 \nonumber \\ &&+
 \left.\sum\limits_{j_{1},j_{2}=0,1,\ldots,\infty}^{j_{1}\neq j_{2}}\hat{b}_{2j_{1}+1}\hat{b}_{2j_{2}+1}S_{25}(j_{1},j_{2})\right ],
     \label{3-10} \end{eqnarray}
  \begin{eqnarray}
 &&\hat{I}_{5}=\frac{b_{0}^{2}}{2L}\left
 [\sum\limits_{l=1,2,\ldots,\infty}\nu(k_{2l})(\hat{b}^{+}_{2l})^{2}(1+S_{22}(l))
 \right.+\nonumber \\&&+
 \sum\limits_{j=0,1,\ldots,\infty}\nu(k_{2j+1})(\hat{b}^{+}_{2j+1})^{2}(1+S_{23}(j))+\nonumber
 \\ &&+
 \sum\limits_{l_{1},l_{2}=1,2,\ldots,\infty}^{l_{1}\neq
 l_{2}}\hat{b}^{+}_{2l_{1}}\hat{b}^{+}_{2l_{2}}S_{24}(l_{1},l_{2})+\nonumber
 \\&&
 +\left.\sum\limits_{j_{1},j_{2}=0,1,\ldots,\infty}^{j_{1}\neq j_{2}}\hat{b}^{+}_{2j_{1}+1}\hat{b}^{+}_{2j_{2}+1}S_{25}(j_{1},j_{2})\right ],
     \label{3-11} \end{eqnarray}
   \begin{eqnarray}
 &&\hat{I}_{6}=\frac{(N-\hat{\tilde{N}})n_{0}\nu(0)}{2}+\nonumber
 \\&&+
 \sum\limits_{l=1,2,\ldots,\infty}n_{0}\nu(k_{2l})\hat{b}^{+}_{2l}\hat{b}_{2l}(1+S_{22}(l)+\tilde{S}_{26}(l))+
 \nonumber \\ &&+
 \sum\limits_{j=0,1,\ldots,\infty}n_{0}\nu(k_{2j+1})\hat{b}^{+}_{2j+1}\hat{b}_{2j+1}\times\nonumber \\
 &&\times(1+S_{23}(j)+\tilde{S}_{27}(j))+\nonumber
 \\ &&+
 \sum\limits_{l_{1},l_{2}=1,2,\ldots,\infty}^{l_{1}\neq
 l_{2}}n_{0}\hat{b}^{+}_{2l_{1}}\hat{b}_{2l_{2}}[S_{24}(l_{1},l_{2})+S_{28}(l_{1},l_{2})]+
 \nonumber \\ &&+
 \sum\limits_{j_{1},j_{2}=0,1,\ldots,\infty}^{j_{1}\neq j_{2}}n_{0}\hat{b}^{+}_{2j_{1}+1}\hat{b}_{2j_{2}+1}
 \times\nonumber \\&&\times [S_{25}(j_{1},j_{2})+S_{29}(j_{1},j_{2})],
     \label{3-12} \end{eqnarray}
 where we denote
\begin{equation}
N-\hat{\tilde{N}}\equiv
\sum\limits_{l=1,2,\ldots,\infty}\hat{b}^{+}_{2l}\hat{b}_{2l}+
 \sum\limits_{j=0,1,\ldots,\infty}\hat{b}^{+}_{2j+1}\hat{b}_{2j+1}.
     \label{3-13a} \end{equation}
Using the relations
$\hat{b}_{l}=e^{-\epsilon_{0}t/i\hbar}\hat{a}_{l}=e^{-\epsilon_{0}t/i\hbar}(\hat{d}_{l}-d^{(0)}_{l})$,
(\ref{9b2}) and the normalization condition
$\sum_{l=1,2,\ldots,\infty}\hat{d}^{+}_{l}\hat{d}_{l}=N$,
 we represent (\ref{3-13a}) in the form
\begin{eqnarray}
&&N-\hat{\tilde{N}}=N-\sum\limits_{j=0,1,\ldots,\infty}|f_{2j+1}|^{2}-\nonumber
\\&& -
\sum\limits_{j=0,1,\ldots,\infty}(\hat{b}^{+}_{2j+1}f_{2j+1}+\hat{b}_{2j+1}f^{*}_{2j+1}).
     \label{3-13} \end{eqnarray}
 It is shown in Appendix B that
$\tilde{S}_{26}(l)=\frac{\nu(0)}{2\nu(k_{2l})}-S_{26}(l)$,
$\tilde{S}_{27}(j)=\frac{\nu(0)}{2\nu(k_{2j+1})}-S_{27}(j)$, and
that the corrections $S_{2j}$ with $j=1,2,\ldots,9$ are negligibly
small at $a\ll L$. Taking this into account, we find from
(\ref{3-1})--(\ref{3-13}):
\begin{eqnarray}
&&\hat{H}= \hat{H}_{lin}+\frac{N_{0}n_{0}\nu(0)}{2}+Nn_{0}\nu(0)
 +\nonumber \\&&+\sum\limits_{j=0,1,\ldots,\infty}\left (K(k_{2j+1})-n_{0}\nu_{0}\right
 )|f_{2j+1}|^{2}+ \nonumber \\
 &&+ \sum\limits_{l=1,2,\ldots,\infty}\left [\left (K(k_{2l})+ n_{0}\nu(k_{2l})\right )\hat{b}^{+}_{2l}\hat{b}_{2l}+
 \right.\nonumber \\&&+\left.\frac{(b_{0}^{*})^{2}}{2L}\nu(k_{2l})\hat{b}^{2}_{2l} +
 \frac{b_{0}^{2}}{2L}\nu(k_{2l})(\hat{b}^{+}_{2l})^{2} \right ]+
 \label{3-14} \\ &&+
 \sum\limits_{j=0,1,\ldots}\left [ \left (
 K(k_{2j+1})+n_{0}\nu(k_{2j+1})\right )\hat{b}^{+}_{2j+1}\hat{b}_{2j+1}
 +\right.\nonumber \\&&+\left.\frac{(b_{0}^{*})^{2}}{2L}\nu(k_{2j+1})\hat{b}^{2}_{2j+1}+
 \frac{b_{0}^{2}}{2L}\nu(k_{2j+1})(\hat{b}^{+}_{2j+1})^{2}\right ],
     \nonumber \end{eqnarray}
\begin{eqnarray}
&& \hat{H}_{lin}=
\sum\limits_{j=0,1,\ldots,\infty}(f_{2j+1}^{*}\hat{b}_{2j+1}+f_{2j+1}\hat{b}^{+}_{2j+1})
\cdot\nonumber \\&&\cdot \left
[K(k_{2j+1})+n_{0}\nu(k_{2j+1})/2-n_{0}\nu_{0}/2 \right ].
      \label{3-15} \end{eqnarray}
The correction term $\hat{H}_{lin}$ is linear in $\hat{b}_{2j+1}$
and $\hat{b}^{+}_{2j+1}$ and can be removed with the help of the
transformation
\begin{eqnarray}
 \hat{b}_{2j+1}=\hat{\tilde{b}}_{2j+1}+\beta_{2j+1}, \ \
 \hat{b}^{+}_{2j+1}=\hat{\tilde{b}}^{+}_{2j+1}+\beta^{*}_{2j+1},
      \label{3-16} \end{eqnarray}
\begin{eqnarray}
 \beta_{2j+1}=-f_{2j+1}\frac{2K(k_{2j+1})+n_{0}\nu(k_{2j+1})-n_{0}\nu_{0}}{2K(k_{2j+1})+4n_{0}\nu(k_{2j+1})},
      \label{3-17} \end{eqnarray}
where  $j=0,1,\ldots,\infty$.  The operators
$\hat{\tilde{b}}^{+}_{2j+1}$ and $\hat{\tilde{b}}_{2p+1}$ satisfy
the same commutation relations as the operators $\hat{b}^{+}_{2j+1}$
and $\hat{b}_{2p+1}$. Thus, we have found the correction $\delta
\psi_{0}(x,t)$:
\begin{eqnarray}
&&\delta \psi_{0}(x,t)\equiv \langle \hat{\psi}(x,t)\rangle -
\psi_{0}(x,t)  =\nonumber \\&&=
e^{\epsilon_{0}t/i\hbar}\sum\limits_{j=0,1,\ldots,\infty}\beta_{2j+1}
  \sqrt{2/L}\cdot   \sin(k_{2j+1}x).
     \label{dpsi} \end{eqnarray}
In the previous Section the correction $\beta_{2j+1}$ did not arise,
because a more approximate solution was found. Let us substitute the
operators $\hat{b}_{2j+1}$, $\hat{b}^{+}_{2j+1}$ (\ref{3-16}) in
Eqs. (\ref{3-14}), (\ref{3-15}). After some transformations, we
obtain the following total Hamiltonian:
\begin{eqnarray}
&& \hat{H}=\frac{N_{0}n_{0}\nu(0)}{2}+
 (N-N_{0})n_{0}\nu(0)+ \delta E_{0}+
  \nonumber \\
 &&+ \sum\limits_{l=1,2,\ldots,\infty}\left [\left (K(k_{2l})+ n_{0}\nu(k_{2l})\right )\hat{b}^{+}_{2l}\hat{b}_{2l}+
\right.\nonumber \\&&+\left.
\frac{(b_{0}^{*})^{2}}{2L}\nu(k_{2l})\hat{b}^{2}_{2l} +
 \frac{b_{0}^{2}}{2L}\nu(k_{2l})(\hat{b}^{+}_{2l})^{2} \right ]+
 \label{3-19} \\ &&+
 \sum\limits_{j=0,1,\ldots,\infty}\left [ \left (
 K(k_{2j+1})+n_{0}\nu(k_{2j+1})\right )\times\right. \nonumber\\&&\times \hat{\tilde{b}}^{+}_{2j+1}\hat{\tilde{b}}_{2j+1}
 +\frac{(b_{0}^{*})^{2}}{2L}\nu(k_{2j+1})\hat{\tilde{b}}^{2}_{2j+1}+\nonumber\\&&+
 \left.\frac{b_{0}^{2}}{2L}\nu(k_{2j+1})(\hat{\tilde{b}}^{+}_{2j+1})^{2}\right ],
     \nonumber \end{eqnarray}
\begin{eqnarray}
&& \delta E_{0}=N_{0}n_{0}\nu(0)+\nonumber
\\&&+\sum\limits_{j=0,1,\ldots,\infty}|f_{2j+1}|^{2}
 \left (K(k_{2j+1})-n_{0}\nu_{0}\right ) -\label{3-20} \\ &&-
 \sum\limits_{j=0,1,\ldots,\infty}|\beta_{2j+1}|^{2}\left (K(k_{2j+1})+2n_{0}\nu(k_{2j+1})\right ).
      \nonumber \end{eqnarray}
 With the help of transforms (\ref{25a}),
 where we use the operators $\hat{b}^{+}_{2l}, \hat{b}_{2l}$ for even $p$ and the operators $\hat{\tilde{b}}^{+}_{2j+1},
\hat{\tilde{b}}_{2j+1}$ for odd $p,$ Hamiltonian (\ref{3-19}) is reduced to
the diagonal form:
\begin{eqnarray}
 \hat{H}=E_{0}+
 \sum\limits_{l=1,2,\ldots,\infty}E_{b}(k_{l})\hat{\xi}^{+}_{l}\hat{\xi}_{l},
      \label{3-21} \end{eqnarray}
\begin{eqnarray}
&&E_{0}= \frac{N_{0}n_{0}\nu(0)}{2} +(N-N_{0})n_{0}\nu(0) +\delta
E_{0}+\nonumber \\&&+\frac{1}{2}\sum\limits_{l=1,2,\ldots,\infty}
 \left (E_{b}(k_{l})-K(k_{l})-n_{0}\nu(k_{l})\right ).
      \label{3-22} \end{eqnarray}
We remark that the operators $\hat{\xi}^{+}_{l}$, $\hat{\xi}_{p}$
(\ref{25}) (with replacements $\hat{b}_{2j+1}\rightarrow
\hat{\tilde{b}}_{2j+1}$, $\hat{b}^{+}_{2j+1}\rightarrow
\hat{\tilde{b}}^{+}_{2j+1}$ or without them) satisfy the commutation
relations (\ref{12}) for Bose operators.

According to solution (\ref{3-21}), (\ref{3-22}),  the system can be
considered as a gas of noninteracting quasiparticles with the energy
$E_{b}(k)$, and the number of quasiparticles can vary. Our solution
differs from the Bogoliubov one by that the quasiparticles are
standing waves rather than traveling ones, which is natural, and by
the term $\delta E_{0}$. It is shown in Appendix B that $\delta
E_{0}$ is  negligible provided that the coupling is weak and the
interaction radius is less or comparable with the interatomic
distance. Under these conditions, our solutions $ E(k)$ and $E_{0}$
coincide with Bogoliubov ones for a periodic system.

We note that if the interaction radius tends to zero, then the
energy levels $E_{b}(k)$ with $k=\pi/L, 2\pi/L, \ldots $ coincide
with the corresponding levels \cite{mt2015,mtjpa2018} of a system of
point bosons, which is described via the Bethe ansatz.

Since the excitations are standing waves, index  $l$ in
 (\ref{3-21}), (\ref{3-22}) takes only positive values ($1,2,3,\ldots$). As $k_{l}$ varies with the step ${\sss
\triangle}k=\pi/L$, the value of the sum $\sum_{l}f(k_{l})$ for zero
BCs is the same as for periodic BCs, for which $l=\pm 1, \pm 2,\pm
3,\ldots$ and ${\sss \triangle}k=2\pi/L$ \cite{bog1947}. Therefore,
the values of thermodynamic quantities under zero and periodic BCs
are identical (the same result was obtained for point bosons
\cite{mt2015}).

We remark  also that the replacement by a c-number was made twice
(rough and fine tunings): first, we separated $d^{(0)}_{2j+1}$ from
$\hat{d}_{2j+1}$ and, second, $\beta_{2j+1}$ from $\hat{b}_{2j+1}$.
The latter is equivalent to the separation of an additional c-number
$\beta_{2j+1}e^{\epsilon_{0}t/i\hbar}$ from $\hat{d}_{2j+1}$: that
is, $d_{2j+1}=d^{(0)}_{2j+1}+\beta_{2j+1}e^{\epsilon_{0}t/i\hbar}$.
If the $(2j+1)$-state  is macroscopically occupied, then
$|\beta_{2j+1}|\ll |d^{(0)}_{2j+1}|$.

Note that in the Hamiltonian we  neglected small corrections of the
kind $\hat{a}^{3}_{j}$, $\hat{a}^{4}_{j}$, which leads to the
absence of the interaction between quasiparticles. Therefore, the
model describes only such states of the system, for which the
interaction of quasiparticles is inessential, i.e., the states with
a not large number of quasiparticles. The criterion of applicability
of the method (see Section 5 below) imposes the stronger restriction
on the number of quasiparticles.

It is also worth noting that, in the Bogoliubov method after the
replacement $\hat{d}_{0}\rightarrow d_{0}$, the exact Hamiltonian
(\ref{3-1})-- (\ref{3-3}) no longer commute with the operator
$\hat{N}=\Sigma_{j}\hat{d}^{+}_{j}\hat{d}_{j}$ of the total number
of particles. Therefore, the number of particles $N$ is not
conserved. Several close modifications of the Bogoliubov method, in
which the number of particles is conserved, were proposed
\cite{girardeau1959,gardiner1997,girardeau1998,zagrebnov2007}. In
our method, the replacement (\ref{9b}) \textit{does not } cause the
violation of the equality $[\hat{H},\hat{N}]= 0$. However, for the
diagonal Hamiltonian (\ref{3-21}), we have $[\hat{H},\hat{N}] \neq
0$. The law of conservation of $N$ is broken due to the neglect of
corrections of the orders $\vartheta^{3}$ and $\vartheta^{4}$ in the
exact $\hat{H}$. It is necessary to modify the model so that the law
of conservation of $N$ be satisfied. On the other hand, in order
that $\langle \hat{\psi}(x,t)\rangle$ exist and be nonzero, one
needs to break the invariance of the Hamiltonian relative to the
transformation $\hat{\psi}\rightarrow e^{i\alpha}\hat{\psi}$
\cite{bogquasi}, which causes the violation of the law of
conservation of $N$ (at least, formally).

 \section{A criterion of applicability of the method}
To find the condition of applicability of the method, we will need
to calculate the ``anomalous'' averages $\langle
\hat{d}^{+}_{l}\hat{d}_{j}\rangle|_{l\neq j}$ and $\langle
\hat{\psi}(x,t)\rangle$. The symmetry-based reasoning
\cite{bogquasi} implies that the law of conservation of the number
of particles leads to $\langle \hat{\psi}(x,t)\rangle=0$. However,
the Bogoliubov model gives a nonzero value of $\langle
\hat{\psi}(x,t)\rangle$. This contradiction can be removed with the
help of the introduction of a negligible correction with a certain
structure into the Hamiltonian (the method of quasiaverages
\cite{bogquasi}). In this case, the law of conservation of the
number of particles is formally broken, the average $\langle
\hat{\psi}(x,t)\rangle$ can be nonzero, and the method of
quasiaverages allows one to find it. However, in order to calculate
$\langle \hat{\psi}(x,t)\rangle,$ there is no need to use the method
of quasiaverages. It is sufficient to use in formula (\ref{srT}) the
quasiparticle representation \cite{bog1949}: We express
$\hat{d}^{+}_{j}$ and $\hat{d}_{j}$ in the operator $\hat{A}$ in
(\ref{srT}) through the operators $\hat{\xi}^{+}_{j}$ and
$\hat{\xi}_{j}$. Then we construct the collection
$\{\Psi_{p}(x_{1},\ldots,x_{N})\}$ from the wave function of the
ground state $\Psi_{0}$, the wave functions of states with one
quasiparticle ($C_{l}\hat{\xi}^{+}_{l}\Psi_{0}$), two quasiparticles
($C_{lj}\hat{\xi}^{+}_{l}\hat{\xi}^{+}_{j}\Psi_{0}$), and so on
($C_{l}$ and $C_{lj}$ are normalization factors). These functions
are the eigenfunctions of Hamiltonian (\ref{3-21}). The functions
$\Psi_{p}$ of a Bose gas have the same structure \cite{bz1955}. It
is known that the Schr\"{o}dinger equation with given BCs has a set
of solutions, which form a complete collection
$\{\Psi_{p}(x_{1},\ldots,x_{N})\}$. It is clear that lowest states
in this collection should coincide with the above-constructed ones.
The energy levels in (\ref{srT}) are described by the formula
\begin{equation}
E_{p}=E_{0}+\sum\limits_{l=1,2,\ldots,\infty}n_{l}E_{b}(k_{l}),
      \label{4-0} \end{equation}
where the occupation numbers take values $n_{l}=0,1,2,\ldots,\infty$
for all $l$'s. In such way, we can find any averages. In this case,
the system of interacting particles with a fixed $N$ is described as
the ideal gas of quasiparticles, whose number varies.

Let us find the criterion of applicability of the method. The
equations
\begin{equation}
\sum_{j=1,2,\ldots,\infty}\hat{N}_{j}=\hat{N}, \quad
\hat{N}_{j}=\hat{d}^{+}_{j}\hat{d}_{j}
 \label{4-01} \end{equation}
yield
\begin{equation}
\sum_{j=1,2,\ldots,\infty}\frac{\langle\hat{d}^{+}_{j}\hat{d}_{j}\rangle}{N}=1,
 \label{4-02} \end{equation}
where $\langle\hat{d}^{+}_{j}\hat{d}_{j}\rangle
=\langle\hat{N}_{j}\rangle$ is the average number of atoms in the
state $\sqrt{2/L}\cdot\sin(k_{j}x)$. For $j=2l$ we have $\langle
\hat{d}_{2l}^{+}\hat{d}_{2l}\rangle=\langle
\hat{b}_{2l}^{+}\hat{b}_{2l}\rangle$. In order to determine $\langle
\hat{b}_{2l}^{+}\hat{b}_{2l}\rangle$ by formula (\ref{srT}), we
express $\hat{b}_{2l}^{+}$, $\hat{b}_{2l}$ in terms of
$\hat{\xi}^{+}_{2l}$, $\hat{\xi}_{2l}$ according to (\ref{25a}) and
take into account that $\langle \hat{\xi}^{+}_{p}\rangle = \langle
\hat{\xi}_{p}\rangle =0$. We obtain
\begin{equation}
\langle \hat{b}_{2l}^{+}\hat{b}_{2l}\rangle=\frac{\langle
\hat{n}_{2l}\rangle +|\Lambda_{2l}|^{2}(\langle \hat{n}_{2l}\rangle
+1)}{1-|\Lambda_{2l}|^{2}},
      \label{4-2} \end{equation}
\begin{equation}
\frac{|\Lambda_{p}|^{2}}{1-|\Lambda_{p}|^{2}}=
\frac{(n_{0}\nu(k_{p}))^{2}(2E_{b}(k_{p}))^{-1}}{E_{b}(k_{p})+K(k_{p})+n_{0}\nu(k_{p})}.
      \label{4-10} \end{equation}
In the approximation of free quasiparticles, the average number of
quasiparticles with the quasimomentum $k=k_{l}$ is determined by the
Bose distribution \cite{bog1949,bose1924,pethick2008}:
\begin{equation}
\langle\hat{n}_{l}\rangle =
\langle\hat{\xi}^{+}_{l}\hat{\xi}_{l}\rangle =
(e^{E_{b}(k_{l})/k_{B}T}-1)^{-1}.
      \label{4-29} \end{equation}
In this case, $\langle\hat{n}_{l}\rangle|_{T=0} =0$. Formula
(\ref{4-2}) with $\langle\hat{n}_{2l}\rangle =0$ follows also from
(\ref{sr0}), (\ref{25a}).

With regard for (\ref{9b}) and (\ref{3-16}), we obtain
\begin{eqnarray}
&&\langle \hat{d}_{2j+1}^{+}\hat{d}_{2j+1}\rangle=\langle
(\hat{\tilde{b}}_{2j+1}^{+}+\beta_{2j+1}^{*}+f^{*}_{2j+1})\cdot\nonumber
\\&&\cdot(\hat{\tilde{b}}_{2j+1}+\beta_{2j+1}+f_{2j+1}) \rangle.
      \label{4-6} \end{eqnarray}
By expressing the operators $\hat{\tilde{b}}_{2j+1}^{+}$,
$\hat{\tilde{b}}_{2j+1}$ through $\hat{\xi}^{+}_{2j+1}$,
$\hat{\xi}_{2j+1}$ according to (\ref{25a}), we find
\begin{equation}
\langle \hat{d}_{2j+1}^{+}\hat{d}_{2j+1}\rangle=G_{2j+1}+\langle
\hat{\tilde{b}}_{2j+1}^{+}\hat{\tilde{b}}_{2j+1} \rangle,
      \label{4-7} \end{equation}
\begin{equation}
G_{2j+1}=(\beta_{2j+1}^{*}+f^{*}_{2j+1})(\beta_{2j+1}+f_{2j+1}),
      \label{4-8} \end{equation}
\begin{equation}
\langle \hat{\tilde{b}}_{2j+1}^{+}\hat{\tilde{b}}_{2j+1}
\rangle=\frac{\langle \hat{n}_{2j+1}\rangle
+|\Lambda_{2j+1}|^{2}(\langle \hat{n}_{2j+1}\rangle
+1)}{1-|\Lambda_{2j+1}|^{2}}.
      \label{4-7a} \end{equation}
The above formulae yield
\begin{eqnarray}
&&1-\frac{\tilde{N}_{0}}{N}=\frac{1}{N}\cdot\nonumber
\\&&\cdot\sum_{j=1,2,\ldots,\infty}\frac{(n_{0}\nu(k_{j}))^{2}}{2E_{b}(k_{j})[E_{b}(k_{j})+K(k_{j})+n_{0}\nu(k_{j})]}+
\nonumber \\
&&+\frac{1}{N}\sum_{j=1,2,\ldots,\infty}\frac{K(k_{j})+n_{0}\nu(k_{j})}{[e^{E_{b}(k_{j})/k_{B}T}-1]E_{b}(k_{j})},
      \label{4-11} \end{eqnarray}
where $\tilde{N}_{0}$ is the number of atoms in the effective
condensate $\langle \hat{\psi}(x,t)\rangle$:
\begin{eqnarray}
&&\tilde{N}_{0}=\int\limits_{0}^{L}|\langle
\hat{\psi}(x,t)\rangle|^{2}dx= \label{4-31}
\\&&=\sum_{j=0,\ldots,\infty}\langle\hat{d}^{+}_{2j+1}(t)\rangle\langle\hat{d}_{2j+1}(t)\rangle
=\sum_{j=0,\ldots,\infty}G_{2j+1}.
      \nonumber \end{eqnarray}
Here,
$\langle\hat{d}_{2j+1}(t)\rangle=e^{\epsilon_{0}t/i\hbar}[f_{2j+1}+\beta_{2j+1}]$.
According to the above analysis, the effective condensate is
described by the formula
\begin{eqnarray}
\langle
\hat{\psi}(x,t)\rangle=\sum\limits_{j=0,1,\ldots,\infty}\langle\hat{d}_{2j+1}(t)\rangle\sqrt{\frac{2}{L}}\cdot\sin(k_{2j+1}x).
      \label{4-30} \end{eqnarray}

For simplicity we consider the point potential
$U(|x_{j}-x_{l}|)=2c\delta(x_{j}-x_{l})$ (i.e. $\nu(k)=2c$) and set
$n=\frac{N}{L}$, $\frac{K(k)}{n_{0}2c}=\frac{k^{2}}{2n_{0}n
\gamma}$, where $\gamma=\frac{2mc}{\hbar^{2}n}$ coincides with
$\gamma$ by Lieb-Liniger \cite{ll1963}.  Then
\begin{eqnarray}
f_{j}+\beta_{j}=f_{j}\frac{4\gamma n_{0}n}{k_{j}^{2}+4\gamma
n_{0}n},
 \label{4-13a} \end{eqnarray}
\begin{eqnarray}
G_{j}=|f_{j}|^{2}\frac{(4\gamma n_{0}n)^{2}}{(k_{j}^{2}+4\gamma
n_{0}n)^{2}}=\frac{8N_{0}}{\pi^{2}j^{2}}\frac{1}{\left
[y_{j}^{2}/4+1 \right ]^{2}},
 \label{4-13g} \end{eqnarray}
\begin{eqnarray}
\tilde{N}_{0}\equiv\sum_{j=0,1,\ldots,\infty}G_{2j+1}=N_{0}\left
(1-\frac{1.5}{\pi\sqrt{\Gamma }}\right ),
      \label{4-32} \end{eqnarray}
 \begin{eqnarray}
&&\chi_{j}\equiv \frac{
(n_{0}\nu(k_{j}))^{2}}{2E_{b}(k_{j})[E_{b}(k_{j})+K(k_{j})+n_{0}\nu(k_{j})]}
=\nonumber
\\&&=\frac{2}{\sqrt{y_{j}^{4}+4y_{j}^{2}}\left (\sqrt{y_{j}^{4}+4y_{j}^{2}}+y_{j}^{2}+2 \right )},
      \label{4-12} \end{eqnarray}
\begin{eqnarray}
&&\omega_{j}\equiv
\frac{K(k_{j})+n_{0}\nu(k_{j})}{[e^{E_{b}(k_{j})/k_{B}T}-1]E_{b}(k_{j})}
=\nonumber
\\ &&=\frac{y_{j}^{2}+2}{[e^{\sqrt{y_{j}^{4}+4y_{j}^{2}}/\tilde{T}}-1]\sqrt{y_{j}^{4}+4y^{2}_{j}}},
      \label{4-13} \end{eqnarray}
\begin{eqnarray}
 \sum\limits_{j=1,2,\ldots,\infty}\chi_{j}=q_{0}(\Gamma)\frac{\sqrt{\Gamma}\ln{\Gamma}}{4},
 \label{4-14a} \end{eqnarray}
\begin{eqnarray}
\sum\limits_{j=1,2,\ldots,\infty}\omega_{j}=q_{T}(\Gamma,\tilde{T})\cdot
0.82\tilde{T}\Gamma,
      \label{4-14} \end{eqnarray}
where $y_{j}=\frac{k_{j}}{\sqrt{\gamma
nn_{0}}}=\frac{j}{\sqrt{\Gamma}}$,
$\tilde{T}=\frac{k_{B}T}{cn_{0}}$, $\Gamma=\frac{\gamma
NN_{0}}{\pi^{2}}$. The equality (\ref{4-32}) holds at $\Gamma \gsim
1$.

Relations (\ref{4-11})--(\ref{4-14}) yield
\begin{eqnarray}
1-\frac{\tilde{N}_{0}}{N}=
\frac{q_{0}(\Gamma)\sqrt{\Gamma}}{4N}\ln{\Gamma}+q_{T}(\Gamma,\tilde{T})\frac{0.82\tilde{T}\Gamma}{N},
      \label{4-17} \end{eqnarray}
where $q_{0}(\Gamma)\approx 1$ at $\Gamma\gg 1$, and $q_{T}\approx
1$ at $\Gamma\gg 1$, $\tilde{T}\gg \Gamma^{-1/2}$.  The condition
$\tilde{T}\gg \Gamma^{-1/2}$ follows from $E_{b}(k_{1})\ll k_{B}T$
(nonzero temperature has a meaning, if the system is in the
equilibrium state; the equilibrium is possible for a large number of
quasiparticles; the last is satisfied for $E_{b}(k_{1})\ll k_{B}T$).

Under periodic BCs the solution \cite{bog1947} yields formula
(\ref{4-17}) with the replacement $\tilde{N}_{0}\rightarrow N_{0}$
and with the parameters $q_{0}\approx 0.9$, $q_{T}\approx 0.5$ (at
$\Gamma\gg 1$, $\tilde{T}\gg \Gamma^{-1/2}$).

We have used the approximation $\hat{\vartheta}\ll \langle
\hat{\psi}(x,t)\rangle$, which holds for $N-\tilde{N}_{0}\ll N$. The
condition $N-\tilde{N}_{0}\ll N$ implies that almost all atoms are
in the effective condensate $\langle \hat{\psi}(x,t)\rangle$. From
relation (\ref{4-17}), we obtain the \textit{criterion} of
applicability of the method:
\begin{equation}
N-\tilde{N}_{0}\ll N \  \Leftrightarrow \
0<\frac{\sqrt{\gamma}}{2\pi}\ln{\frac{N\sqrt{\gamma}}{\pi}}+0.08\gamma
N\tilde{T}\ll 1.
      \label{4-18} \end{equation}
At $N=\infty $, $T=0$ inequalities (\ref{4-18}) are broken for any
finite $\gamma$. From (\ref{4-32}), (\ref{4-17}) we get
$q_{0}\approx 1$, $\tilde{N}_{0}=N_{0}\approx N\left
(\frac{2\pi}{\sqrt{\gamma}\ln{N}}\right)^{2}\ll N$. Such value of
$\tilde{N}_{0}$ can be considered macroscopic at $\sqrt{\gamma}\lsim
2\pi$ (for the system with finite $N$ we consider the $j$-th state
to be macroscopically occupied, if $\langle \hat{N}_{j}\rangle\gsim
N/\Theta$, where $\Theta=\ln^{2}{N}$). If $N=\infty $  and $T> 0$,
then (\ref{4-17}) implies that $\tilde{N}_{0}$ is microscopic.  It
was shown in \cite{frag2018} that $\tilde{N}_{0}$ is close to the
number of atoms in the unique condensate. Such results agree with
the conclusion of the well-known works \cite{kk1967,hohenberg1967}
that the existence of a condensate in an infinite 1D system at $T>0$
is impossible.

At finite $N$ and $T=0$ inequalities (\ref{4-18}) hold provided that
$\gamma$ is small. At finite $N$  and  $T>0$ relation (\ref{4-18})
yields $\gamma N\tilde{T}\ll 12$, and the condition $\tilde{T}\gg
\Gamma^{-1/2}$ gives $\tilde{T}\gg \pi/(\sqrt{\gamma} N)$. These
inequalities are compatible at $\gamma < 10^{-3}$. Additionally, the
relation
$\frac{\sqrt{\gamma}}{2\pi}\ln{\frac{N\sqrt{\gamma}}{\pi}}\ll 1$
should hold. This shows that, in a finite system, criterion
(\ref{4-18}) holds even at $T>0$, if $\gamma < 10^{-3}$ and the
temperature is small: $\tilde{T}\ll \frac{12}{\gamma N}$, i.e.,
$k_{B}T \ll \frac{6\hbar^{2}n^{2}}{mN}$. Interestingly, the last
inequality contains no interaction constant $c$. At first sight, the
condition $k_{B}T \ll \frac{6\hbar^{2}n^{2}}{mN}$ differs radically
from the criterion of Bose condensation for the ideal Bose gas in a
1D trap: $k_{B}T\lsim \frac{N\hbar \omega}{\ln{(2N)}}$
\cite{kd1996}, where $\omega$ is the trap frequency. But this is not
the case. Zero BCs are similar to a trap with the frequency
$\hbar\omega= \frac{\hbar^{2}k_{min}^{2}}{2m}$, where $
k_{min}=\frac{\pi}{L}$. Therefore, the criterion $k_{B}T\lsim
\frac{N\hbar \omega}{\ln{(2N)}}$ leads to $k_{B}T \lsim
\frac{\pi^{2}\hbar^{2}n^{2}}{2mN\ln{(2N)}}$, which is close to
$k_{B}T \ll \frac{6\hbar^{2}n^{2}}{mN}$.

Thus, our method can be used for the description of a finite  system
at low temperature and with weak coupling.

\section{Single-particle density matrix and the quasicondensate}

For the completeness, we now find the density matrix:
\begin{eqnarray}
&&F_{1}(x,x^{\prime})=\langle
\hat{\psi}^{+}(x^{\prime},t)\hat{\psi}(x,t) \rangle =\nonumber \\
&&=
\sum\limits_{j_{1}j_{2}=1,2,\ldots,\infty}g^{*}_{2j_{1}-1}(x^{\prime})g_{2j_{2}-1}(x)\cdot
\nonumber \\ &&\cdot\left
[d^{(0)*}_{2j_{1}-1}d^{(0)}_{2j_{2}-1}+d^{(0)*}_{2j_{1}-1}\langle\hat{a}_{2j_{2}-1}\rangle +\right.\nonumber \\
&&+\left.
d^{(0)}_{2j_{2}-1}\langle\hat{a}^{+}_{2j_{1}-1}\rangle+\langle
\hat{a}^{+}_{2j_{1}-1}\hat{a}_{2j_{2}-1}\rangle\right ]+\nonumber
\\&&
+\sum\limits_{l_{1}l_{2}=1,2,\ldots,\infty}g^{*}_{2l_{1}}(x^{\prime})g_{2l_{2}}(x)\langle
\hat{a}^{+}_{2l_{1}}\hat{a}_{2l_{2}}\rangle +\nonumber \\
&&+\sum\limits_{j_{1}l_{2}=1,2,\ldots,\infty}g^{*}_{2j_{1}-1}(x^{\prime})g_{2l_{2}}(x)\cdot
\label{5-10}\\&&\cdot\left [
d^{(0)*}_{2j_{1}-1}\langle\hat{a}_{2l_{2}}\rangle+\langle
\hat{a}^{+}_{2j_{1}-1}\hat{a}_{2l_{2}}\rangle\right ]+ \nonumber \\
&&+\sum\limits_{l_{1}j_{2}=1,2,\ldots,\infty}g^{*}_{2l_{1}}(x^{\prime})g_{2j_{2}-1}(x)\cdot
\nonumber\\&&\cdot\left [
d^{(0)}_{2j_{2}-1}\langle\hat{a}^{+}_{2l_{1}}\rangle+\langle
\hat{a}^{+}_{2l_{1}}\hat{a}_{2j_{2}-1}\rangle\right ].
     \nonumber \end{eqnarray}
We take into account that
$\hat{a}_{p}(t)=e^{\epsilon_{0}t/i\hbar}\hat{b}_{p}$, $
\hat{b}_{2j-1}=\hat{\tilde{b}}_{2j-1}+\beta_{2j-1}$,
$d^{(0)}_{2j-1}=e^{\epsilon_{0}t/i\hbar}f_{2j-1}$. Let us express
the operators $\hat{\tilde{b}}_{2j-1}, \hat{b}_{2j}$ in terms of
$\hat{\xi}^{+}_{p}, \hat{\xi}_{p}$ with the help of relations
(\ref{25a}). We can verify that $\langle \hat{b}_{2j}
\rangle=\langle \hat{\tilde{b}}_{2j-1} \rangle=0$, $\langle
\hat{b}^{+}_{2j} \rangle=\langle \hat{\tilde{b}}^{+}_{2j-1}
\rangle=0$, $\langle
\hat{\tilde{b}}^{+}_{2j_{1}-1}\hat{\tilde{b}}_{2j_{2}-1}
\rangle=\delta_{j_{1},j_{2}}\langle
\hat{\tilde{b}}^{+}_{2j_{1}-1}\hat{\tilde{b}}_{2j_{1}-1} \rangle$,
$\langle \hat{b}^{+}_{2l_{1}}\hat{b}_{2l_{2}}
\rangle=\delta_{l_{1},l_{2}}\langle
\hat{b}^{+}_{2l_{1}}\hat{b}_{2l_{1}}\rangle$, $\langle
\hat{\tilde{b}}^{+}_{2j_{1}-1}\hat{b}_{2l_{2}} \rangle=\langle
\hat{b}^{+}_{2l_{1}}\hat{\tilde{b}}_{2j_{2}-1} \rangle=0$. Then
relation (\ref{5-10}) yields
\begin{eqnarray}
&&F_{1}(x,x^{\prime})=\langle
\hat{\psi}(x^{\prime},t)\rangle^{*}\cdot\langle
\hat{\psi}(x,t)\rangle +\nonumber \\&&
+\sum\limits_{l=1,2,\ldots,\infty}g^{*}_{2l}(x^{\prime})g_{2l}(x)\langle
\hat{b}^{+}_{2l}\hat{b}_{2l}\rangle +\nonumber \\ &+&
\sum\limits_{j=1,2,\ldots,\infty}g^{*}_{2j-1}(x^{\prime})g_{2j-1}(x)\langle
\hat{\tilde{b}}^{+}_{2j-1}\hat{\tilde{b}}_{2j-1}\rangle.
     \label{5-11} \end{eqnarray}
With regard for the formula $g_{l}(x)=\sqrt{2/L}\cdot\sin{(k_{l}x)}$
and relations (\ref{4-2}), (\ref{4-10}), (\ref{4-29}), (\ref{4-7a}),
for the point interaction ($\nu(k)=2c$), relation (\ref{5-11})
yields
\begin{eqnarray}
&&F_{1}(x,x^{\prime})=f_{0}^{*}(x^{\prime})f_{0}(x)+\nonumber\\&&
+\frac{2}{L}\sum\limits_{l=1,2,\ldots,\infty}
\frac{\sin{(k_{l}x^{\prime})}\sin{(k_{l}x)}}{\sqrt{y_{l}^{4}+4y_{l}^{2}}}\cdot
\nonumber \\  &\cdot &\left
(\frac{2}{\sqrt{y_{l}^{4}+4y^{2}_{l}}+y^{2}_{l}+2}+\frac{y^{2}_{l}+2}{e^{\frac{\sqrt{y_{l}^{4}+4y_{l}^{2}}}{\tilde{T}}}-1}
\right ),
     \label{5-14} \end{eqnarray}
\begin{equation}
f_{0}(x)=
\frac{4\sqrt{n_{0}}}{\pi}\sum\limits_{j=1,2,\ldots,\infty}\frac{\sin{(k_{2j-1}x)}}{2j-1}
\frac{4}{y^{2}_{2j-1}+4},
     \label{5-15} \end{equation}
where $y_{l}=l/\sqrt{\Gamma}$. The total concentration of the gas is
given by the formula
\begin{eqnarray}
n(x)=F_{1}(x,x).
     \label{5-13} \end{eqnarray}

\begin{figure}
\vskip1mm
\includegraphics[width=\column]{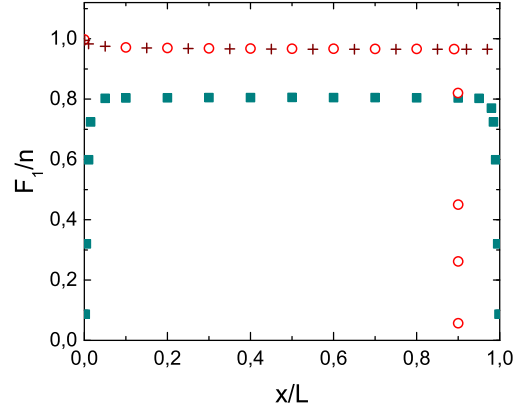}
\vskip-3mm\caption{[Color online]  Concentration
$n(x)/n=F_{1}(x,x)/n$ (squares, $N=10^{4}$,  $\gamma=0.0001$) and
the density matrix $F_{1}(x_{1},x_{1}+x)/n$ (open circles,
$x_{1}=0.1L$, $N=10^{5}$, $\Gamma=10^{6}$, $\gamma\approx 0.00102$)
under zero BCs determined numerically by formulae (\ref{5-14}),
(\ref{5-15}) at $T=0$. Crosses show solution (\ref{5-16}) for the
density matrix under periodic BCs. In order to separate the curves,
the values of the concentration $n(x)$ are multiplied by $0.8$.  }
\end{figure}

In Fig. 1 we show the concentration $n(x)$ and the density matrix
$F_{1}(x_{1},x_{1}+x)$ (as the function of $x$),  determined
numerically from (\ref{5-14})--(\ref{5-13}) for $T=0$. In this case,
we obtained $N_{0}$ from (\ref{4-32}), (\ref{4-17}). As is seen, the
concentration is constant inside the system, decreases by
approaching the boundary, and turns to zero on the boundary. The
width of a band near the boundary, where the concentration varies,
is $\delta L \approx L/\sqrt{\Gamma}= \pi/\sqrt{\gamma nn_{0}}$. The
density matrix $F_{1}(x_{1},x_{1}+x)$ depends on $x$ and is almost
independent of $x_{1}$, if the point $x_{1}$ is at a distance $\gsim
\delta L$ from the boundaries. It is seen from Fig. 1 that
$F_{1}(x_{1},x_{1}+x)|_{T=0}$ (\ref{5-14}) is very close to the
solution for a periodic system at $T=0$, which reads
\cite{pethick2008,mtjltp2016,schwartz1977,haldane1981,petrov2004,popov1972,berkovich1989,mora2003,olshanii2009}
\begin{equation}
F_{1}(x_{1},x_{1}+x)|_{|x|\gsim \delta L} \approx n\left
(\frac{\delta L\cdot f_{2}(|x|)}{|x|}\right )^{\sqrt{\gamma}/2\pi}.
     \label{5-16} \end{equation}

The available literature gives several close values for $f_{2}(x)$:
$1/\pi$ \cite{schwartz1977,pethick2008,haldane1981,petrov2004},
$0.33$ \cite{popov1972,berkovich1989,mora2003}, and
$\frac{1}{3}[0.98+2(x/L)^{2}]$ \cite{mtjltp2016}. In Fig. 1, we
present function (\ref{5-16}) with
$f_{2}(x)=\frac{1}{3}[0.98+2(x/L)^{2}]$. Such $f_{2}(x)$ leads to a
slightly better agreement than $f_{2}(x)=1/\pi; 0.33$.  It is seen
that, as $x$ varies from $0$ to $L-x_{1},$ the function
$F_{1}(x_{1},x_{1}+x)$ is changed only by $3\%$ at $\gamma\approx
0.00102$, $|x|\gsim \delta L$. For $\gamma=0.0001,$ the function
$F_{1}(x_{1},x_{1}+x)$ varies only by $0.5\%$, as $x$ increases. If
$F_{1}(x_{1},x_{1}+x)$ decreases (as the function of $x$) by a power
law, the term ``quasicondensate'' \cite{bouchoule2009}  is usually
used instead of ``condensate''. But our solution
$F_{1}(x_{1},x_{1}+x)$ decreases, as $x$ increases, \textit{very
slowly} for any parameters $\gamma$ and $N$ satisfying criterion
(\ref{4-18}). That is,  we have the \textit{true condensate} for
such parameters. Such regime was found previously \cite{mtjltp2016}.
For the applicability of our method it is insignificant whether
$\langle \hat{\psi}(x,t)\rangle$ corresponds to the condensate or
quasicondensate. It is important that the system be finite.

Note that the curve $F_{1}(x_{1},x_{1}+x)$ (\ref{5-14}) (circles in
Fig. 1) drops to zero at $x= 0.9L$. This is due to the fact that
Fig. 1 is plotted for $x_{1}=0.1L$. So, for $x= 0.9L$ the coordinate
$x_{2}$ turns out to be on the boundary: $x_{2}\equiv x_{1}+x=L$.

Our analysis shows that, under zero BCs, the power law (\ref{5-16})
holds for all $x_{1}\in [\delta L, L-\delta L]$, $x\in [\delta L,
L-x_{1}-\delta L]$. In  works by Cazalilla
\cite{cazalilla2002,cazalilla2004}, $F_{1}(x,x^{\prime})$ was also
calculated for a 1D system of interacting bosons with zero BCs.
According to those results, $F_{1}(x,x^{\prime})$ decreases by a
power law, as $|x-x^{\prime}|$ increases, only at the points near
the vessel center. The deviation from a power law for the remaining
points is related, apparently, to the harmonic-fluid approximation
used in \cite{cazalilla2002,cazalilla2004}.

We remark that the function $F_{1}(0,x)|_{T=0}$ was also calculated
for a periodic system of point bosons at intermediate values of
$\gamma$ and finite $N$ \cite{astra2006,caux2007}.

By neglecting the terms $\hat{\vartheta}$ and $ \delta \psi_{0}$ in
the formula $\hat{\psi} = \psi_{0}(x,t) +\delta \psi_{0}(x,t)+
\hat{\vartheta}(x,t)$,  we obtain
\begin{eqnarray}
&&F_{1}(x,x^{\prime})= \psi^{*}_{0}(x^{\prime},t)\psi_{0}(x,t) =
\nonumber \\&&=\left [ \begin{array}{ccc}
    n_{0},  & \   \mbox{if} \ x,x^{\prime}\in ]0,L[,   & \\
    0,  & \mbox{if} \ x=0; L \ \mbox{or} \ x^{\prime}=0; L. &
     \label{5-18} \end{array}     \right.   \end{eqnarray}
In this case, we lost both the smooth decrease in $F_{1}$ near the
boundaries and the power decrease by law (\ref{5-16}).

Using formulae (\ref{5-14}) and (\ref{5-15}), we have found
numerically the density matrix for $T>0$. At $\delta L \lsim
x_{1}\lsim L-\delta L$, $10 \delta L \lsim x\lsim L-10 \delta L$,
$\frac{\pi}{\sqrt{\gamma}N}\ll \tilde{T} \lsim \frac{4}{\gamma N},$
the solution takes the form
\begin{equation}
F_{1}(x_{1},x_{1}+x)|_{T>0}\approx  C_{T}(x_{1})n e^{-0.5\gamma
n\tilde{T}xx_{1}L^{-1}},
     \label{5-0T} \end{equation}
where $C_{T}(x_{1})$ is the normalization factor depending on
$x_{1}$. If criterion (\ref{4-18}) ($\gamma N\tilde{T}\lsim 1$) is
satisfied, the function $F_{1}(x_{1},x_{1}+x)$ (\ref{5-0T})
decreases weakly, as $x$ increases. This corresponds to the true
condensate. At  $0<\tilde{T}\lsim \frac{\pi}{\sqrt{\gamma}N}$ the
solution is intermediate between  (\ref{5-16}) and (\ref{5-0T}). We
did not find it, since such temperatures are not quite physical (at
such $\tilde{T}$ the system contains only several quasiparticles,
and the thermal equilibrium is impossible). We note that solution
(\ref{5-0T}) \textit{differs} significantly from the solution for a
periodic system, which is independent of $x_{1}$
\cite{popov1972,schwartz1977,mora2003}:
\begin{eqnarray}
&&F_{1}(x_{1},x_{1}+x)|_{T>0} \approx \nonumber
\\&&\approx n\left (\frac{\delta L\cdot
f_{2}(|x|)}{|x|}\right )^{\sqrt{\gamma}/2\pi}e^{-0.25\gamma
n_{0}\tilde{T}|x|}.
     \label{5-pT} \end{eqnarray}

Let us write the wave function of the effective condensate in the
form $\langle \hat{\psi}(x,t)\rangle
=\sqrt{n_{c}(x)}e^{i\alpha(x,t)}$, $n_{c}(x)=f^{2}_{0}(x)$. The
numerical solution indicates that, at $\gamma \ll 1$, $T=0$, the
condensate concentration $n_{c}(x)$ is very close to the
 total concentration $n(x)$: $n_{c}(x)=
(1-|\kappa(x)|)n(x)$, where $|\kappa(x)|\ll 1$. In particular, for
$N=10^{4}$, $\gamma=0.0001$ we have $\kappa(x)= 0.001 - 0.007$.

We remarked above that the concentration $n(x)|_{T=0}$ is nonuniform
near the boundaries in the layer of thickness $\delta L \approx
\pi/\sqrt{\gamma nn_{0}}$. For $\gamma \lsim N^{-2}$ we obtain
$\delta L \gsim L$: the concentration is nonuniform in the whole
system. This corresponds to the regime of almost free particles. Our
model does not work for it, since the approximation
$\hat{\vartheta}\ll \langle \hat{\psi}(x,t)\rangle$ is violated. At
$\gamma \lsim N^{-2}$ the correction $\delta E_{0}$ (\ref{3-20})
becomes large (see Appendix B). The value of $E_{0}$ for this regime
was found for point bosons, described via the Bethe ansatz
\cite{mt2015,batchelor2005}.

 \section{Summary}
We have generalized Bogoliubov method and have constructed a
description of low-lying levels of a one-dimensional system of
weakly interacting bosons under zero boundary conditions. Two points
are significant for the method: (i) the representation
$\hat{\psi}(x,t) = \langle \hat{\psi}(x,t)\rangle
+\hat{\vartheta}(x,t)$ and (ii) separation of the principal part in
integrals. We emphasize that the method requires the existence of
the order parameter $\langle \hat{\psi}(x,t)\rangle$, rather than a
condensate (determined on the basis of a diagonal expansion of the
density matrix). Sometimes, the average $\langle
\hat{\psi}(x,t)\rangle$ coincides with a condensate \cite{bog1947};
sometimes, $\langle \hat{\psi}(x,t)\rangle$ is only close to the
condensate \cite{frag2018}. Apparently, it is possible that the
condensate (quasicondensate) is absent, but $\langle
\hat{\psi}(x,t)\rangle \neq 0,$ and the method is valid. Though we
were not faced with such case yet.

In the subsequent work \cite{frag2018} we diagonalized the density
matrix (\ref{5-14}). It turned out that the average $\langle
\hat{\psi}(x,t)\rangle$ does not quite coincide with the true
quasicondensate defined on the basis of expansion (\ref{n2}).
Moreover, the quasicondensate is fragmented, and its structure
depends on the boundary conditions.

In Appendix А we argue that the Bogoliubov quasiparticles are
collective excitations, though they look as one-particle
excitations.

It is important that our solutions for $E_{0}$ and $E(k)$ coincide
with that ones obtained by the \textit{exactly solvable approach},
based of the Bethe ansatz \cite{gaudin1971,mt2015,mtjpa2018}. In
addition, our solution for the density matrix
$F_{1}(x_{1},x_{2})|_{T=0}$ coincides with the solution obtained
within other methods for periodic BCs
\cite{pethick2008,mtjltp2016,schwartz1977,haldane1981,petrov2004,popov1972,berkovich1989,mora2003,olshanii2009}.
This clearly shows that the Bogoliubov approximation is quite
accurate for a finite 1D Bose system with weak coupling.
Interestingly, for the point interaction, the Bogoliubov solutions
$E_{0}$ and $E(k)$ are approximately valid also at $\gamma \sim 1$
even in the limit $N\rightarrow \infty$
\cite{ll1963,mt2015,mtjpa2018}, which contradicts criterion
(\ref{4-18}). This means that the region of applicability of the
Bogoliubov solutions is much wider than the region of applicability
of the Bogoliubov method.

We have found that the bulk properties ($E_{0}$ and $E(k)$) of the
system with boundaries are the same as for a periodic system. This
is understandable in view of the definition
\begin{eqnarray}
&&E_{0}=\int dx_{1}\ldots dx_{N}\Psi_{0}^{*}\left
[-\frac{\hbar^2}{2m}\sum\limits_{j=1}^{N}\frac{\partial^2}{\partial
x_{j}^2}+ \right.\nonumber
\\&&+\left.\sum\limits_{j<l}U(|x_{j}-x_{l}|)\right ]\Psi_{0}.
      \label{4-22} \end{eqnarray}
In order that $E_{0}$ depend on the boundaries, it is necessary that
the roughness of a landscape of the wave function
$\Psi_{0}(x_{1},\ldots, x_{N})$ or the probability of overlapping of
two adjacent atoms be dependent on the boundaries. However, both
these properties have a local nature and, apparently, should not
depend on the remote boundaries. If such dependence would exist, we
would have a nontrivial effect. In order that such effect be, it is
necessary that a solution with a substantially smaller energy than
$E_{0}$ obtained above (see Sec. 4)  exist. However, our ground
state is characterized by the uniform effective condensate $\langle
\hat{\psi}(x,t)\rangle$, which contains almost all atoms. We see no
state, which would possess a smaller energy. Therefore, we may
conclude with a high degree of confidence that the boundaries do not
affect the bulk properties of a system of weakly interacting bosons.
Though it would be more interesting, if such influence would exist.

\vskip3mm \textit{The author is grateful to Yu. Shtanov for the
valuable discussion. The present work is partially supported by the
National Academy of Sciences of Ukraine (project No.~0116U003191).}

 \section{Appendix A. Remarks on the Bogoliubov method}
According to the Bogoliubov transformations (\ref{25}), the operator
of creation of a quasiparticle $\hat{\xi}^{+}_{p}$ is the sum of the
operators of creation ($\hat{b}^{+}_{p}$) and annihilation
($\hat{b}_{p}$) of \textit{one} particle. From whence many authors
made conclusion about the single-particle nature of Bogoliubov
quasiparticles. However, the solutions by Feynman \cite{fey1954} and
by Bogoliubov--Zubarev \cite{bz1955} clearly show  that \textit{a
quasiparticle is a collective oscillation of the whole gas}
(condensate and above-condensate atoms). Let us consider this point.

First, one-particle excitations are impossible in one dimension,
since the moving atom will necessarily collide with neighbor atom
\cite{giamarchi2003}. But the Bogoliubov solutions in 1D, 2D, and 3D
cases are similar. Hence, the Bogoliubov quasiparticles in 1D, 2D,
and 3D should be collective and should coincide with quasiparticles
in \cite{fey1954,bz1955}. The solutions for $E_{0}$ and $E(k)$
obtained in the Bogoliubov \cite{bog1947} and collective
\cite{bz1955} approaches coincide.

Next. The method of secondary quantization is based on that the
$N$-particle Bose-symmetric wave function of the system can be
presented as the expansion \cite{bog1949}
\begin{eqnarray}
&&\Psi(x_{1},\ldots,x_{N})=
\sum\limits_{\{n_{f}\}}C(n_{f_{1}},\ldots,n_{f_{N}})\cdot\nonumber
\\&&\cdot \psi_{\{n_{f}\}} (x_{1},\ldots,x_{N}),
     \label{7-1} \end{eqnarray}
\begin{equation}
\psi_{\{n_{f}\}}
(x_{1},\ldots,x_{N})=\tilde{c}_{\{n_{f_{j}}\}}\sum\limits_{P}P\varphi_{f_{1}}(x_{1})\ldots\varphi_{f_{N}}(x_{N}).
     \label{7-2} \end{equation}
Here, $\{n_{f}\}=(n_{f_{1}},\ldots,n_{f_{N}})$, $n_{f_{1}}+\ldots
+n_{f_{N}}=N$, the functions $\varphi_{l}(x)$ form a complete
orthonormalized collection,  the number $n_{f_{j}}=0,1,2,\ldots $
indicates the number of identical functions $\varphi_{f_{j}}(x)$ in
expansion (\ref{7-2}) (each index $f_{j}$ runs the same values as
$l$), and $P$ means all possible permutations. Based on formulae
(\ref{7-1}), (\ref{7-2}), and
\begin{equation}
\hat{\psi}(x) = \sum\limits_{l}\hat{b}_{l}\varphi_{l}(x),
     \label{7-3} \end{equation}
one can obtain the basic formulae of the method: (\ref{2}) and
(\ref{3-1})--(\ref{3-3}) \cite{bog1949}. We now pass from
$\hat{b}^{+}_{p}$, $\hat{b}_{p}$ to the operators of creation and
annihilation of quasiparticles $\hat{\xi}^{+}_{p}, \hat{\xi}_{p}$.
In this case, we should consider the basis functions
$\tilde{\varphi}_{l}(x)$ from the equality $\hat{\psi}(x) =
\sum\limits_{l}\hat{\xi}_{l}\tilde{\varphi}_{l}(x)$ instead of the
functions $\varphi_{l}(x)$. Each eigenfunction $\Psi_{\{n_{f}\}} $
of the Hamiltonian can be represented in the form
$C(\hat{\xi}^{+}_{l_{1}})^{n_{l_{1}}} \ldots
(\hat{\xi}^{+}_{l_{p}})^{n_{l_{p}}}\Psi_{0}$, where $n_{l}$ is the
number of quasiparticles corresponding to index $l$. These numbers
$n_{l}$ coincide with the numbers $n_{l}$ in relation (\ref{4-0})
and do \textit{not} coincide with the numbers $n_{l}$ in expansion
(\ref{7-2}) executed in the single-particle basis
$\{\tilde{\varphi}_{l}(x)\}$ (in the case of coincidence, we would
get the wave function of the ground state
$\Psi_{0}(x_{1},\ldots,x_{N})=const$, since
$n_{f_{1}}=n_{f_{2}}=\ldots = n_{f_{N}}=0$ for the state without
quasiparticles; but the function
$\Psi_{0}(x_{1},\ldots,x_{N})=const$ is not a solution of the
Schr\"{o}dinger equation with an interaction;  moreover, the
requirement $n_{f_{1}}+\ldots +n_{f_{N}}=N$ is violated).
\textit{This means that a quasiparticle is not a single-particle
structure,} despite the single-particle form of formulae (\ref{25}).
We see also that the Bogoliubov method does not allow one to find
the explicit form of the eigenfunctions
$\Psi_{\{n_{f}\}}(x_{1},\ldots,x_{N})$. However, the diagonalization
of the Hamiltonian enables one to find all observable parameters of
the system: lowest energy levels $E_{j}$ (\ref{4-0}), concentration
$n(x)$ (\ref{5-13}), correlation functions, and thermodynamic
parameters.

The experiment \cite{ketterle2002} seems to be consistent with the
single-particle interpretation of formulae (\ref{25}), according to
which $\hat{b}^{+}_{p}$ creates one atom.  Note that the formula
$\hat{b}^{+}_{p}|N_{p}=j\rangle = \sqrt{j+1}|N_{p}=j+1\rangle$ used
in \cite{ketterle2002} is valid for the ideal gas with a variable
number of particles and gives zero anomalous averages $\langle
\hat{b}^{+}_{\textbf{k}}\hat{b}^{+}_{-\textbf{k}} \rangle, \langle
\hat{b}_{\textbf{k}}\hat{b}_{-\textbf{k}} \rangle $. But in the
Bogoliubov model the averages $\langle
\hat{b}^{+}_{\textbf{k}}\hat{b}^{+}_{-\textbf{k}} \rangle, \langle
\hat{b}_{\textbf{k}}\hat{b}_{-\textbf{k}} \rangle $ are nonzero and
play the important role \cite{bog1947,frag2018}. In the Bogoliubov
approach, the state with one quasiparticle is described by the wave
function
$\hat{\xi}^{+}_{p}\Psi_{0}=(1-|\Lambda_{p}|^{2})^{-1/2}(\hat{b}^{+}_{p}-\Lambda^{*}_{p}\hat{b}_{p})\Psi_{0}$,
which is a superposition of the state with $N+1$ particles and the
state with $N-1$ particles ($\Psi_{0}$ is a state with $N$
particles). In contrast, in the models \cite{fey1954,bz1955} the
wave functions for the ground state and states with one
quasiparticle are eigenfunctions of the Schr\"{o}dinger equation
with the same fixed number of particles $N$. We assume that such
contradiction is due to the limitation of the language of operators
$\hat{b}^{+}_{p}$, $\hat{b}_{p}$: some properties expressed in this
language become distorted.  It arose apparently because the method
of secondary quantization is based on a formalism with a variable
number of particles $N$, but it is used for the calculation of the
energy eigenvalues of the Schr\"{o}dinger equation with fixed $N$.
Despite this difficulty, the Bogoliubov method gives results for the
observable quantities in agreement with other approaches.

We note that the many-particle nature  of excitations in  the
Bogoliubov model is hidden \cite{ketterle2002} in the values of the
coefficients $u_{p}=(1-|\Lambda_{p}|^{2})^{-1/2}$ and
$v_{p}=-\Lambda^{*}_{p}(1-|\Lambda_{p}|^{2})^{-1/2}$ from
(\ref{25}): at a small quasimomentum  $k=\pi p/L$ we have
$u_{p}\approx |v_{p}|\gg 1$ (excitation involves many atoms); for
large
 $k$ we get $u_{p}\simeq 1, v_{p}\simeq 0$ (excitation is mainly connected with one atom).

We remark also that, in the calculation of the average $\langle
\hat{A}(x,t)\rangle$, one needs to use the canonical ensemble (for a
system with fixed $N$) or the grand canonical ensemble (for a system
with variable $N$). In a system with zero BCs, the number of
particles must be conserved. Therefore, in formula (\ref{srT}) we
based ourselves on the canonical ensemble.  Bogoliubov
\cite{bog1947} used the canonical ensemble as well (it is clearly
seen from the analysis in \cite{bog1949}).

\section{Appendix B. Corrections $S_{j}$ and $\delta E_{0}$}
1) Below, we list the formulae and estimates for the functions
$S_{j}(x)$ from Sections 3 and 4. The numerical calculations are
carried out for the potential
\begin{equation}
U(|x-x^{\prime}|)=\frac{c_{0}}{a}e^{-|x-x^{\prime}|/a},
      \label{6-1} \end{equation}
which is characterized by the interaction radius $a$ and  the
Fourier-transform $\nu(k_{j})=\frac{2c_{0}(1-e^{-L/a}\cos(\pi
j))}{1+(k_{j}a)^{2}}$, $k_{j}=\pi j/L$. We consider the interaction
radius to be small ($a\ll L$) and set $e^{-L/a}=0$. For each
function $S_{j},$ we give the exact formula and a numerical estimate
made for potential (\ref{6-1}). The results are the following.
\begin{eqnarray}
&&S_{1}(x)=\frac{2}{L}\sum\limits_{j=1,3,5,\ldots,\infty}\frac{\sin(k_{j}x)}{k_{j}}\left
( \frac{\nu(k_{j})}{\nu(0)}-1 \right )\approx \nonumber \\&&\approx
\frac{1}{\pi}\int\limits_{0}^{\infty}dy\frac{\sin(yx/a)}{y}\left (
\frac{\nu(y/a)}{\nu(0)}-1 \right ).
      \label{6-2} \end{eqnarray}
Here, for potential (\ref{6-1})  we have $S_{1}(x=0, L)=-0.5$ on the
boundaries. As the distance from boundaries increases, $|S_{1}|$
decreases. For $x\in ]5a,L-5a[$, we have $|S_{1}|\lsim a/L$. That
is, the correction $S_{1}$ is not small near the boundaries.
Therefore, solution (\ref{20}) of Eq. (\ref{13}) is incorrect near
the boundaries. We neglected this fact in Section 3, since the
inaccuracy of the solution near the boundaries should not affect the
bulk properties of the system.
\begin{eqnarray}
&&S_{2}(l)=S_{22}(l)=\sum\limits_{j=1,3,5,\ldots,\infty}\frac{2}{\pi^{2}}\left
( \frac{\nu(k_{j})}{\nu(k_{2l})}-1 \right )\cdot\nonumber
\\&&\cdot\left ( \frac{1}{2l-j}+ \frac{1}{2l+j} \right )^{2},
\label{6-3a} \end{eqnarray}
\begin{eqnarray}
0<S_{2}\lsim 80 l^{2}\left (\frac{a}{L}\right )^{3},
      \label{6-3} \end{eqnarray}
 \begin{eqnarray}
&&S_{3}(l_{1},l_{2})|_{l_{1}\neq
l_{2}}=S_{24}(l_{1},l_{2})=\frac{2}{\pi^{2}}\sum\limits_{j=1,3,\ldots,\infty}
\nu(k_{j})\cdot\nonumber\\ &&\cdot \left
(\frac{1}{2l_{1}-j}+\frac{1}{2l_{1}+j} \right )\left
(\frac{1}{2l_{2}-j}+\frac{1}{2l_{2}+j} \right ), \nonumber \\&&
0<S_{3} \lsim \frac{3}{\pi^{2}}c_{0}l_{1}l_{2}\left
(\frac{8a}{L}\right )^{3},
      \label{6-4} \end{eqnarray}
\begin{eqnarray}
&&S_{4}(j)=S_{23}(j)=\sum\limits_{l=0,1,\ldots,\infty}
\frac{\nu(k_{2l})-\nu(k_{2j+1})}{\nu(k_{2j+1})}\cdot
\nonumber\\&&\cdot \frac{2q(l)}{\pi^{2}} \cdot \left
(\frac{1}{2j+1-2l}+\frac{1}{2j+1+2l} \right )^{2}=\nonumber
\\ &&= 4\frac{a^{2}}{L^{2}} \left
(1+\right. \label{6-5}
\\&&\left.+\sum\limits_{l=1,2,\ldots}\frac{2}{1+(2\pi
la/L)^{2}}\frac{1}{1-(2l/(2j+1))^{2}}\right ),
      \nonumber \end{eqnarray}
\begin{eqnarray}
0<S_{4}\lsim 20(2j+1)^{2}\left (\frac{a}{L}\right )^{3},
      \label{6-6} \end{eqnarray}
  \begin{eqnarray}
&&S_{5}(j_{1},j_{2})|_{j_{1}\neq
j_{2}}=S_{25}(j_{1},j_{2})=\nonumber\\&&=\frac{2}{\pi^{2}}\sum\limits_{l=0,1,2,\ldots,\infty}
\nu(k_{2l})q(l)\cdot \nonumber \\ &&\cdot  \left
(\frac{1}{2j_{1}+1-2l}+\frac{1}{2j_{1}+1+2l} \right )\cdot\nonumber
\\&&\cdot\left (\frac{1}{2j_{2}+1-2l}+\frac{1}{2j_{2}+1+2l} \right ),
      \label{6-7} \end{eqnarray}
 \begin{eqnarray}
0<S_{5} \lsim \frac{6}{\pi^{2}}c_{0}(2j_{1}+1)(2j_{2}+1)\left
(\frac{4a}{L}\right )^{3},
      \label{6-8} \end{eqnarray}
\begin{equation}
S_{21}=\frac{4}{\pi^{2}}\sum\limits_{j=1,3,5,\ldots,\infty}
\frac{\nu(k_{j})-\nu(0)}{\nu(0)j^{2}}\simeq -\frac{a}{L},
      \label{6-21} \end{equation}
\begin{eqnarray}
&&\tilde{S}_{26}(l)=\frac{2}{\pi^{2}}\sum\limits_{j=1,3,5,\ldots,\infty}
\frac{\nu(k_{j})}{j\nu(k_{2l})}\cdot\label{6-26a} \\&&\cdot\left
(\frac{2}{j}-\frac{1}{j-4l}-\frac{1}{j+4l} \right )=
\frac{\nu(0)}{2\nu(k_{2l})}-S_{26}(l), \nonumber \end{eqnarray}
\begin{eqnarray}
40 l^{2}\left (\frac{a}{L}\right )^{3} \lsim S_{26}(l)\lsim 160
l^{2}\left (\frac{a}{L}\right )^{3},
      \label{6-26} \end{eqnarray}
 \begin{eqnarray}
&&\tilde{S}_{27}(j)=\frac{2}{\pi^{2}}\sum\limits_{j_{0}=1,3,5,\ldots,\infty}
\frac{\nu(k_{j_{0}})}{j_{0}\nu(k_{2j+1})}\cdot\nonumber
\\&&\cdot\left
(\frac{2}{j_{0}}-\frac{1}{j_{0}-4j-2}-\frac{1}{j_{0}+4j+2} \right )=
 \nonumber \\ &=& \frac{\nu(0)}{2\nu(k_{2j+1})}-S_{27}(j),
 \label{6-27}   \end{eqnarray}
\begin{eqnarray}
(2j+1)^{2}\left (\frac{a}{L}\right )^{3} \lsim 0.1S_{27}(j)\lsim
(4j+2)^{2}\left (\frac{a}{L}\right )^{3},
     \nonumber \end{eqnarray}
\begin{eqnarray}
&&S_{28}(l_{1},l_{2})=\frac{2}{\pi^{2}}\sum\limits_{j=1,3,5,\ldots,\infty}
\frac{\nu(k_{j})}{j}\cdot\nonumber \\&&\cdot\left
(\frac{1}{j-2l_{1}+2l_{2}}+\frac{1}{j+2l_{1}-2l_{2}} - \right.\label{6-28a} \\
&&-\left.\frac{1}{j-2l_{1}-2l_{2}} -\frac{1}{j+2l_{1}+2l_{2}}\right
), \nonumber \end{eqnarray}
\begin{eqnarray}
-320 c_{0}l_{1}l_{2} \left (\frac{a}{L}\right )^{3}\lsim S_{28} <0,
      \label{6-28} \end{eqnarray}
 \begin{eqnarray}
&&S_{29}(j_{1},j_{2})=\frac{2}{\pi^{2}}\sum\limits_{j=1,3,5,\ldots,\infty}
\frac{\nu(k_{j})}{j}\cdot\nonumber \\&&\cdot \left
(\frac{1}{j-2j_{1}+2j_{2}}+\frac{1}{j+2j_{1}-2j_{2}} - \right.\label{6-29a} \\
&&-\left.\frac{1}{j-2j_{1}-2j_{2}-2}
-\frac{1}{j+2j_{1}+2j_{2}+2}\right ), \nonumber \end{eqnarray}
\begin{eqnarray}
 -c_{0}(2j_{1}+1)(2j_{2}+1)
\left (\frac{a}{L}\right )^{3}\lsim \frac{S_{29}}{80} <0.
      \label{6-29} \end{eqnarray}
The values of all $S_{j\geq 2}$ (except for $S_{26}$ and $S_{27}$),
dependent on one or two parameters ($j,l,j_{1},l_{1}\ldots$),
increase in modulus with each parameter, if the parameter is at most
$L/a$. If the parameter exceeds $L/a$, then $|S_{j}|$ almost does
not increase. The values of $S_{26}$ and $S_{27}$ increase with the
parameter according to (\ref{6-26}), (\ref{6-27}) for any values of
the parameter.

In the above formulae, we use the function
\begin{equation}
 q(l) =
\left [ \begin{array}{ccc}
    1/2  & \   l=0,   & \\
    1  & \ l=1,2,\ldots,\infty. &
\label{6-30} \end{array} \right. \end{equation}

The following formulae are also useful for the calculation of sums
($p$, $l$, $j$ are integers):
\begin{equation}
 \sum\limits_{j=\pm 1,\pm 3,
\pm 5,\ldots} \frac{1}{(2l-j)(2p-j)}=\frac{\pi^{2}}{4}\delta_{p,l},
      \nonumber \end{equation}
\begin{eqnarray}
&&\sum\limits_{j=1,3,5,\ldots}\left ( \frac{1}{2l-j}+ \frac{1}{2l+j}
\right )\cdot\nonumber \\&&\cdot\left ( \frac{1}{2p-j}+
\frac{1}{2p+j} \right )=\frac{\pi^{2}}{4}(\delta_{p,l}-
\delta_{p,-l}),
      \nonumber \end{eqnarray}
 \begin{equation}
\sum\limits_{p=0,\pm 1,\pm 2,\ldots}
\frac{1}{(2l+1-2p)(2j+1-2p)}=\frac{\pi^{2}}{4}\delta_{j,l},
      \nonumber \end{equation}
 \begin{eqnarray}
&& \sum\limits_{p=0, 1, 2,\ldots}\left
(\frac{q(p)}{2l+1-2p}+\frac{q(p)}{2l+1+2p}\right )\cdot\nonumber
\\&& \cdot \left (\frac{1}{2j+1-2p}+\frac{1}{2j+1+2p}\right )=\nonumber \\&&=
\frac{\pi^{2}}{4}(\delta_{j,l}-\delta_{2l+1,-2j-1}).
      \nonumber \end{eqnarray}

2) Let us find the quantity $\delta E_{0}$ (\ref{3-20}). With the use of the relation
 $1+\frac{1}{3^{2}}+\frac{1}{5^{2}}+\ldots = \frac{\pi^{2}}{8}$,
 formula (\ref{3-20}) can be written in the form
\begin{eqnarray}
&& \delta
E_{0}=\sum\limits_{j=1,3,5,\ldots,\infty}|f_{j}|^{2}\cdot\nonumber
\\&&\cdot \frac{4K(k_{j})(n_{0}\nu(k_{j})+n_{0}\nu_{0})
-(n_{0}\nu(k_{j})-n_{0}\nu_{0})^{2}}{4K(k_{j})+8n_{0}\nu(k_{j})}=
\nonumber\\&&=\frac{2n_{0}^{3/2}\nu_{0}}{\sqrt{\gamma
n}}I(\Upsilon),
 \label{6-34} \end{eqnarray}
\begin{eqnarray}
 I(\Upsilon)=\int\limits_{x_{min}}^{\infty}\frac{dx}{\pi}\frac{1+\Xi(x)-
 [\Upsilon x\Xi(x)]^{2}/8}{x^{2}+\Xi(x)},
      \label{6-35} \end{eqnarray}
where  $\Upsilon=4\gamma N_{0}Na^{2}/L^{2}$, $\Xi(x)=(1+\Upsilon
x^{2})^{-1}$, $\gamma=\frac{2mc_{0}}{\hbar^{2}n}$,
$x_{min}=\frac{1}{2\sqrt{\Gamma}}$, and potential (\ref{6-1}) is
used. The weak coupling implies that $\gamma \ll 1$. We find
numerically $I(\Upsilon\lsim 1,x_{min} \ll 1)\approx 1$. For
uncharged particles we have $a\sim L/N$, $\Upsilon\sim \gamma$. This
leads to estimates $I\approx 1$ and $\delta E_{0}\sim
2n_{0}\nu_{0}/\sqrt{\gamma}\sim E_{0}/(N\sqrt{\gamma})$  for
$N^{-2}\ll \gamma \lsim 1$. That is, $\delta E_{0}$ is negligible
for the weak coupling (but for $\gamma \gg N^{-2}$), which is
consistent with the solutions for point bosons
\cite{gaudin1971,mt2015,batchelor2005}. For $\gamma \lsim N^{-2}$
the correction $\delta E_{0}$ is not small:  $\delta E_{0} \sim
Nn_{0}\nu_{0} \sim E_{0}$.

\vspace*{-5mm} \rezume{%
М.Д. Томченко} {НИЖНІ ЕНЕРГЕТИЧНІ РІВНІ ОДНОВИМІРНОГО СЛАБКО
ВЗАЄМОДІЮЧОГО БОЗЕ-ГАЗУ \\ З НУЛЬОВИМИ МЕЖОВИМИ УМОВАМИ} {Ми
діагоналізували вторинно квантований гамільтоніан одновимірного
бозе-газу для відштовхувального міжатомного потенціалу загального
вигляду та нульових межових умов. При малій константі зв'язку
розв'язки для енергії основного стану $E_{0}$ та закону дисперсії
$E(k)$ співпадають з відомими  розв'язками для періодичної системи.
При цьому одночастинкова матриця густини $F_{1}(x,x^{\prime})$ є
близькою до розв'язку для періодичної системи, якщо $T=0$, та
помітно відрізняється від останнього при $T>0$.  Також ми отримали,
що хвильова функція $\langle \hat{\psi}(x,t) \rangle$ ефективного
конденсату близька до константи $\sqrt{N_{0}/L}$ всередині системи
та обертається на нуль на межах (тут $N_{0}$ --- число атомів у
конденсаті, $L$ --- розмір системи). Ми знайшли критерій
застосовності методу, згідно з яким метод працює для скінченної
системи з малою константою зв'язку (слабка взаємодія або велика
концентрація) та дуже малою температурою.}


\begin{thebibliography}{200}
\bibitem {bog1947} N.N.~Bogoliubov, On the theory of superfluidity, J.~Phys. USSR \textbf{11}, 23 (1947).
\bibitem {fey1954}  R. Feynman, Atomic theory of the two-fluid model of liquid helium, Phys. Rev. \textbf{94}, 262 (1954)
[DOI: https://doi.org/10.1103/PhysRev.94.262].
\bibitem {bz1955}  N.N. Bogoliubov,  D.N.~Zubarev, The wave function of the lowest state of a system of
interacting Bose particles, Sov. Phys. JETP \textbf{1}, 83 (1955).
\bibitem {brueck1959}  K. Brueckner,  {\it Theory of Nuclear Structure}
        (Methuen, 1959).
\bibitem {ll1963}  E.H. Lieb, W.~Liniger, Exact analysis of an interacting Bose gas. I. The general solution and the
ground state, Phys. Rev. \textbf{130}, 1605 (1963) [DOI:
https://doi.org/10.1103/PhysRev.130.1605].
\bibitem {zero-liquid}  M.D. Tomchenko, Microstructure of He II in the presence of boundaries, Ukr. J. Phys. \textbf{59}, 123 (2014).
\bibitem {gaudin1971}  M. Gaudin, Boundary energy of a Bose gas in one dimension, Phys. Rev. A \textbf{4}, 386 (1971)
[DOI: https://doi.org/10.1103/PhysRevA.4.386].
\bibitem {mt2015}  M. Tomchenko, Point bosons in a one-dimensional box: the ground state, excitations and
thermodynamics, J.~Phys.~A: Math. Theor. \textbf{48}, 365003 (2015)
[DOI: https://doi.org/10.1088/1751-8113/48/36/365003].
\bibitem {mtjpa2018}  M. Tomchenko, Quasimomentum of an elementary excitation for a system of point
bosons with zero boundary conditions, arXiv:1705.10565
[cond-mat.quant-gas].
\bibitem {cazalilla2002}  M.A. Cazalilla, Low-energy properties of a one-dimensional system
of interacting bosons with boundaries, EPL \textbf{59}, 793 (2002)
[DOI: https://doi.org/10.1209/epl/i2002-00112-5].
\bibitem {cazalilla2004}  M.A. Cazalilla, Bosonizing one-dimensional cold atomic gases,
 J.~Phys.~B: At. Mol. Opt. Phys. \textbf{37}, S1 (2004) [DOI:
https://doi.org/10.1088/0953-4075/37/7/051].
\bibitem {girardeau1959}  M.D. Girardeau, R. Arnowitt, Theory om many-boson systems: pair theory, Phys.
Rev. \textbf{113}, 755 (1959) [DOI:
https://doi.org/10.1103/PhysRev.113.755].
\bibitem {bogquasi} N.N. Bogoliubov, Quasi-Averages in problems of statistical mechanics, Dubna report D-781 (1961) (in Russian);
N.N. Bogoliubov, \textsl{Lectures on Quantum Statistics}, vol. 2:
\textsl{Quasi-Averages} (Gordon and Breach, 1970) [ISBN
0-677-20570-8].
\bibitem {fetter1972}  A.L. Fetter, Nonuniform states of an imperfect Bose gas,
 Ann. Phys. \textbf{70}, 67 (1972) [DOI: https://doi.org/10.1016/0003-4916(72)90330-2].
\bibitem {gardiner1997}  C.W. Gardiner, Particle-number-conserving Bogoliubov method
which demonstrates the validity of the time-dependent
Gross-Pitaevskii equation for a highly condensed Bose gas, Phys.
Rev. A \textbf{56}, 1414 (1997) [DOI:
https://doi.org/10.1103/PhysRevA.56.1414].
\bibitem {girardeau1998}  M.D. Girardeau, Comment on  ``Particle-number-conserving Bogoliubov method
which demonstrates the validity of the time-dependent
Gross-Pitaevskii equation for a highly condensed Bose gas'', Phys.
Rev. A \textbf{58}, 775 (1998) [DOI:
https://doi.org/10.1103/PhysRevA.58.775].
\bibitem {leggett2001}  A.G.~Leggett, Bose-Einstein condensation in the alkali gases: Some
fundamental concepts,  Rev. Mod. Phys. \textbf{73}, 307 (2001) [DOI:
https://doi.org/10.1103/RevModPhys.73.307].
\bibitem {zagrebnov2001} V.A. Zagrebnov, J.-B.~Bru, The Bogoliubov model of weakly
imperfect Bose gas, Phys. Rep. \textbf{350}, 291
          (2001) [DOI: https://doi.org/10.1016/S0370-1573(00)00132-0].
\bibitem {zagrebnov2007} V.A. Zagrebnov, The Bogoliubov theory of weakly
imperfect Bose gas and its modern development, in: N.N. Bogoliubov,
Collection of scientific works in 12 volumes, ed. by A.D. Sukhanov
(Nauka, 2007), v. 8, p. 576 (in Russian) [ISBN 978-5020339422,
978-5-02-035723-5].
\bibitem {rovenchak2007}  A. Rovenchak, Weakly-interacting bosons in a trap within approximate
second quantization approach, J.~Low Temp. Phys. \textbf{148}, 411
(2007) [DOI: https://doi.org/10.1007/s10909-007-9406-x].
\bibitem {rovenchak2016}  A. Rovenchak, Effective Hamiltonian and excitation spectrum of harmonically trapped bosons, Low Temp. Phys. \textbf{42}, 36
(2016) [DOI:  https://doi.org/10.1063/1.4939154].
\bibitem {bzt2015}  V.B. Bobrov, A.G. Zagorodny, S.A. Trigger, Coulomb interaction potential
and Bose-Einstein condensate, Low Temp. Phys. \textbf{41}, 901
(2015). [DOI: https://doi.org/10.1063/1.4936669].
\bibitem {deguchi2013} J. Sato, E.~Kaminishi, T.~Deguchi, Finite-size scaling behavior of
Bose-Einstein condensation in the 1D Bose gas, arXiv:1303.2775
[cond-mat.quant-gas].
\bibitem {streltsov2013} J.~Grond,  A.I.~Streltsov, A.U.J.~Lode, K.~Sakmann,
L.S.~Cederbaum, O.E.~Alon, Excitation spectra of many-body systems
by linear response: General theory and applications to trapped
condensates, Phys. Rev. A \textbf{88}, 023606 (2013) [DOI:
https://doi.org/10.1103/PhysRevA.88.023606].
\bibitem {kk1967}  J.W. Kane, L.P.~Kadanoff, Long-range order in superfluid helium, Phys. Rev. \textbf{155}, 80 (1967)
[DOI: https://doi.org/10.1103/PhysRev.155.80].
\bibitem {hohenberg1967}  P.C. Hohenberg, Existence of long-range order in one and two dimensions,
Phys. Rev.  \textbf{158}, 383 (1967) [DOI:
https://doi.org/10.1103/PhysRev.158.383].
\bibitem {fischer2002} U.R. Fischer, Existence of long-range order for trapping interacting bosons,
 Phys. Rev. Lett. \textbf{89}, 280402 (2002) [DOI: https://doi.org/10.1103/PhysRevLett.89.280402].
\bibitem {bl2007} A.I. Bugrij, V.M.~Loktev, On the theory of Bose-Einstein condensation of
quasiparticles: On the possibility of condensation of ferromagnons
at high temperatures, Low Temp. Phys. \textbf{33}, 37 (2007) [DOI:
https://doi.org/10.1063/1.2409633].
\bibitem {kirzhnits1978} D.A. Kirzhnits, Superconductivity and elementary particles,
Sov. Phys. Usp. \textbf{21}, 470 (1978) [DOI:
10.1070/PU1978v021n05ABEH005556].
\bibitem {griffin2002} A. Griffin, BEC and the new world of coherent matter waves, in {\it Theoretical Physics at the End of the
Twentieth Century}, ed. by Y.~Saint-Aubin and L.~Vinet (Springer,
2002), p. 277 [ISBN 0387953116, 978-0387953113],
arXiv:cond-mat/9911419.
\bibitem {penronz}   O. Penrose, L.~Onsager, Bose-Einstein condensation and liquid helium, Phys. Rev. \textbf{104}, 576 (1956)
[DOI: https://doi.org/10.1103/PhysRev.104.576].
\bibitem {frag2018}  M. Tomchenko, On a fragmented condensate in a uniform Bose system, arXiv:1808.08203 [cond-mat.quant-gas].
\bibitem {bog1949}  N.N. Bogoliubov, \textsl{Lectures on Quantum
Statistics}, vol.~1: \textsl{Quantum Statistics} (Gordon and Breach,
1967) [ISBN  0677200307, 9780677200309].
\bibitem {gross1957} E.P. Gross, Unified theory of interacting bosons,
Phys. Rev. \textbf{106}, 161 (1957) [DOI: https://doi.org/10.1103/PhysRev.106.161].
\bibitem {gross1961} E.P. Gross, Structure of a quantized vortex in boson systems,
Nuovo Cimento \textbf{20}, 454 (1961) [DOI:
https://doi.org/10.1007/BF02731494].
\bibitem {pit1961}  L.P. Pitaevskii, Vortex lines in an imperfect Bose gas, Sov. Phys. JETP \textbf{13}, 451 (1961).
\bibitem {mtpot2014}  M. Tomchenko,  Expansions of the interatomic potential under various
boundary conditions and the transition to the thermodynamic limit,
arXiv:1403.8014 [cond-mat.other].
\bibitem {bose1924}  S.N. Bose, Plancks gesetz und lichtquantenhypothese, Z. Phys. \textbf{26}, 178 (1924).
\bibitem {pethick2008} C.J.~Pethick, H.~Smith, \textit{Bose-Einstein
          Condensation in Dilute Gases} (Cambridge Univ. Press, 2008), Chap. 15 [ISBN 978-0-521-84651-6].
\bibitem {kd1996} W.~Ketterle, N.J.~van Druten, Bose-Einstein condensation of a finite number of particles
trapped in one or three dimensions, Phys. Rev. A \textbf{54}, 656
(1996) [DOI: https://doi.org/10.1103/PhysRevA.54.656].
\bibitem {mtjltp2016}  M. Tomchenko, Bose–Einstein condensation in a one-dimensional system
of interacting bosons, J.~Low Temp. Phys. \textbf{182}, 170 (2016)
[DOI: https://doi.org/10.1007/s10909-015-1435-2].
\bibitem {schwartz1977} M. Schwartz, Off-diagonal long-range behavior of interacting Bose systems,
 Phys. Rev. B \textbf{15}, 1399 (1977) [DOI: https://doi.org/10.1103/PhysRevB.15.1399].
\bibitem {haldane1981} F.D.M. Haldane, Effective harmonic-fluid approach to low-energy properties of
one-dimensional quantum fluids, Phys. Rev. Lett. \textbf{47}, 1840
(1981) [DOI: https://doi.org/10.1103/PhysRevLett.47.1840].
\bibitem {petrov2004} D.S.~Petrov, D.M.~Gangardt,  G.V.~Shlyapnikov,
Low-dimensional trapped gases, J.~Phys. IV France \textbf{116}, 3
(2004).
\bibitem {popov1972} V.N. Popov, On the theory of the superfluidity of two- and one-dimensional bose systems,
 Theor. Math. Phys. \textbf{11}, 565 (1972) [DOI: https://doi.org/10.1007/BF01028373].
\bibitem {berkovich1989} A. Berkovich, G.~Murthy, Time-dependent multipoint correlation functions
of the nonlinear schr\"{o}dinger model, Phys. Lett. A \textbf{142},
121 (1989) [DOI: https://doi.org/10.1016/0375-9601(89)90172-2].
\bibitem {mora2003} C. Mora, Y.~Castin, Extension of Bogoliubov theory to quasicondensates,
Phys. Rev. A \textbf{67}, 053615 (2003) [DOI:
https://doi.org/10.1103/PhysRevA.67.053615].
\bibitem {olshanii2009} V. Dunjko, M.~Olshanii, A Hermite-Pad\'{e} perspective on Gell-Mann--Low renormalization group:
an application to the correlation function of Lieb-Liniger gas,
arXiv:0910.0565 [cond-mat.quant-gas].
\bibitem {bouchoule2009} I. Bouchoule, N.J.~van Druten, C.I.~Westbrook, Atom chips and one-dimensional Bose gases,
 arXiv:0901.3303 [physics.atom-ph].
\bibitem {astra2006} G.E.~Astrakharchik, S.~Giorgini, Correlation functions of a Lieb–Liniger Bose gas,
J.~Phys.~B: At. Mol. Opt. Phys. \textbf{39}, S1 (2006) [DOI:
https://doi.org/10.1088/0953-4075/39/10/S01].
\bibitem {caux2007} J.-S.~Caux, P.~Calabrese, N.A.~Slavnov, One-particle dynamical correlations in the one-dimensional Bose gas, J.~Stat. Mech.
          P01008 (2007) [DOI: https://doi.org/10.1088/1742-5468/2007/01/P01008].
\bibitem {batchelor2005}  M.T. Batchelor, X.W.~Guan, N.~Oelkers,
C.~Lee, The 1D interacting Bose gas in a hard wall box, J.~Phys.~A:
Math. Gen.  \textbf{38}, 7787 (2005) [DOI:
https://doi.org/10.1088/0305-4470/38/36/001].
\bibitem {giamarchi2003}  T. Giamarchi, {\it Quantum Physics in One Dimension}
   (Clarendon Press, 2003), Chap. 1 [ISBN 0-19-852500-1].
\bibitem {ketterle2002} J.M. Vogels, K.~Xu, C.~Raman, J.R.~Abo-Shaeer,  W.~Ketterle,
Experimental observation of the Bogoliubov transformation for a
Bose-Einstein condensed gas,  Phys. Rev. Lett. \textbf{88}, 060402
(2002) [DOI: https://doi.org/10.1103/PhysRevLett.88.060402].

\begin{flushright}
{\footnotesize Received 04.11.18}
\end{flushright}
\end{thebibliography}
       \end{document}